\documentclass[12pt,a4paper]{article}

\usepackage[utf8]{inputenc}
\usepackage[T1]{fontenc}
\usepackage[english]{babel}
\usepackage{mathptmx}
\usepackage[scaled=0.92]{helvet}
\usepackage{courier}
\usepackage{amssymb,amsmath,amsthm}
\usepackage{graphicx}
\usepackage{float}
\usepackage{enumitem}
\usepackage{microtype}
\usepackage{rotating}
\usepackage{color}
\usepackage{multirow}

\usepackage{hyperref}

\newtheorem{thm}{Theorem}[section]
\newtheorem{prop}[thm]{Proposition}
\newtheorem{lemma}[thm]{Lemma}
\newtheorem{cor}[thm]{Corollary}
\newtheorem{@definition}[thm]{Definition}
\newenvironment{defn}{\begin{@definition}\rm}{\end{@definition}}
\newtheorem{@notation}[thm]{Notation}

\newtheorem{@example}[thm]{Example}

\newtheorem{@assumption}[thm]{Assumption}
\newenvironment{assumption}{\begin{@assumption}\rm}{\end{@assumption}}
\newtheorem{@remark}[thm]{Remark}
\newenvironment{remark}{\begin{@remark}\rm}{\end{@remark}}
\newenvironment{prf}[1][Proof]{\begin{proof}[\textsc{#1}]}{\end{proof}}

\numberwithin{equation}{section}

\newcommand{\mal}{\stackrel{\mbox{\tiny$\bullet$}}{}}
\newcommand{\uti}{U}
\newcommand{\auf}{[\![}
\newcommand{\zu}{]\!]}
\newcommand{\const}{\mathrm{cst.}}
\begin{document}
\title{Portfolio Optimization under Small Transaction Costs: a Convex Duality
Approach}
\author{
Jan Kallsen\footnote{
Mathematisches Seminar,
Christian-Albrechts-Universit\"at zu Kiel,
Westring 383,
24098 Kiel, Germany,
(e-mail: kallsen@math.uni-kiel.de).
}
\quad Shen Li\footnote{
Mathematisches Seminar,
Christian-Albrechts-Universit\"at zu Kiel,
Westring 383,
24098 Kiel, Germany,
(e-mail: s.li@math.uni-kiel.de).}
}
\date{}
\maketitle

\begin{abstract}
We consider an investor with constant absolute risk aversion who trades
a risky asset with general Itô dynamics, in the presence of small
proportional transaction costs. Kallsen and Muhle-Karbe \cite{kallsen.muhlekarbe.12}
formally derived the leading-order optimal trading policy and the
associated welfare impact of transaction costs. In the present paper,
we carry out a convex duality approach facilitated by the concept
of shadow price processes in order to verify the main results of \cite{kallsen.muhlekarbe.12}
under well-defined regularity conditions.\\



\end{abstract}

\section{Introduction}
A classical problem of mathematical finance concerns an economic agent 
who invests in a financial market so as to maximize the expected utility of her terminal wealth. 
A possible approach to tackle such problems is based on the dual characterization of admissible 
portfolios with the help of convex analysis. This has been studied mostly in frictionless environments, 
for instance in \cite{karatzas.al.91,schachermayer.99}. In the context of markets with friction, 
Cvitani\'{c} and Karatzas \cite{cvitanic.karatzas.96} extended this approach to problems 
with proportional transaction costs. They rely more or less explicitly on the 
concept of consistent price systems or shadow price processes, which 
allow to translate the original problem into a more tractable frictionless 
one, cf.\ in particular Loewenstein \cite{loewenstein.00} in this context.

In a recent study, Kallsen and Muhle-Karbe \cite{kallsen.muhlekarbe.12} investigate
optimal portfolio choice with respect to exponential utility and small
transaction costs for general Itô processes. They formally derive
a leading-order optimal trading policy and the associated welfare
impact. The purpose of the present study is to rigorously prove the
main statements of \cite{kallsen.muhlekarbe.12} under well-defined regularity
conditions. Our approach resembles that of Henderson \cite{henderson.02},
in the sense that an explicitly known dual control provides us an
upper bound to the optimization problem. Since this bound coincides
to the leading order with the utility of a candidate strategy to the
primal problem, the latter must be approximately optimal.

Starting with \cite{shreve.soner.94, whalley.willmott.97}, 
the problem of optimal investment in the presence of small proportional transaction costs
has been studied extensively. For an account of the literature, 
we refer the reader to \cite{kallsen.muhlekarbe.12,kallsen.muhlekarbe.13}.
Rigorous derivations of leading-order optimal strategies in related
setups are provided e.g.\ in \cite{janecek.shreve.04,bichuch.11} and in
particular \cite{soner.touzi.11,possamai.al.12}. The cited papers
carry a strong analytic flavour. \cite{soner.touzi.11,possamai.al.12}
make use of the deep theory of homogenization and viscosity solutions.
By contrast and as noted above, our more probabilistic approach relies
on dual considerations. In particular, the value function as a key
object in analytical approaches appears only implicitly here. In fact,
even its existence is not obvious if the underlying model fails to be of Markovian
structure.

The paper is organized as follows. The market model is introduced
in Section~\ref{sec:MarketModel}. Subsequently, we state the main results concerning
optimal investment to the leading order. In Section~\ref{sec:Examples}, we present
two classes of examples, namely the Black-Scholes model and a more
general stochastic volatility model. The proofs of the main results
are provided in Section~\ref{sec:Proof}. The appendix contains the derivation of
the frictionless optimizer related to the models of Section~\ref{sec:Examples}.

We generally use the notation as in \cite{js.87}. In particular,
$H\mal Y:=\int_{0}^{\cdot}H_{t}dY_{t}$ stands for the stochastic integral
of $H$ with respect to $Y$.

\section{The market model} \label{sec:MarketModel}
We consider the same setup as in \cite{kallsen.muhlekarbe.12}: 
fixing a finite
time horizon $T\in(0,\infty)$, the financial market consists
of a riskless asset (bond) with price normalized to $1$ and a risky
asset (stock) traded with proportional transaction costs. The stock
price $S$ is modelled by a general Itô process
\[
dS_{t}=b_{t}^{S}dt+\sigma_{t}^{S}dW_{t}
\]
defined on a filtered probability space $(\Omega,\mathcal{F},\mathcal{\mathbf{F}},\mathbf{P})$.
Here, $W$ is a one-dimensional, standard Brownian motion and $b$,
$\sigma$ are predictable processes satisfying
\[
\int_{0}^{T}\left(|b_{t}^{S}|+(\sigma_{t}^{S})^{2}\right)dt<\infty\quad \mbox{a.s.}
\]
Let $\varepsilon\in(0,1)$ denote the relative bid-ask spread,
i.e., an investor has to pay the higher ask\emph{ }price $(1+\varepsilon)S$
but only receives the lower bid\emph{-}price $(1-\varepsilon)S$
for buying and selling the stock, respectively.

\begin{defn} \label{defn:adm}
\begin{enumerate}
	\item A \emph{trading strategy} is an $\mathbb{R}^{2}$-valued predictable process $(\psi^{0},\psi)$ of finite variation, 
	where $\psi_{t}^{0}$ and $\psi_{t}$ denote the number of shares held in the bank account
and in stock at time $t$, respectively.
	\item The \emph{liquidation wealth process }of a trading strategy $(\psi^{0},\psi)$
is defined as
\[
X^{\psi,\varepsilon}:=\psi^{0}+\psi\mathbf{1}_{\{ \psi\geq0\} }
(1-\varepsilon)S+\psi\mathbf{1}_{\{ \psi<0\} }(1+\varepsilon)S.
\]
	\item Writing $\psi=\psi^{\uparrow}-\psi^{\downarrow}$ with increasing
predictable processes $\psi^{\uparrow}$, $\psi^{\downarrow}$ which
do not increase at the same time, a trading strategy $(\psi^{0},\psi)$
is called \emph{self financing} if
\[
d\psi_{t}^{0}=(1-\varepsilon)S_{t}d\psi_{t}^{\downarrow
}-(1+\varepsilon)S_{t}d\psi_{t}^{\uparrow},
\]
cf.\ \cite{kallsen.muhlekarbe.08c}. For given initial value $\psi_{0}^{0}$,
a self-financing trading strategy $(\psi^{0},\psi)$ will
be identified with its second component $\psi$ in the sequel.

\item Given \emph{initial wealth} $(x^{B},x^{S})\in\mathbb{R}^{2}$ in the
bank account and the stock, respectively, a self-financing trading
strategy $(\psi^{0},\psi)$ is said to be \emph{admissible}
for $(x^{B},x^{S})$ and written as
\[
\psi\in\mathcal{A}^{\varepsilon}(x^{B},x^{S})
\]
if $x^{B}=\psi_{0}^{0}$, $x^{S}=\psi_{0}S_{0}$, and if the related
liquidation wealth is bounded from below, i.e.,
\[
X^{\psi,\varepsilon}\geq-K
\]
for some $K\in\mathbb{R}_{+}$.

\end{enumerate} \end{defn}

\begin{remark}\label{rem:self-fin}
The liquidation wealth of a self-financing strategy $(\psi^{0},\psi)$
with $x^{B}=\psi_{0}^{0}$, $x^{S}=\psi_{0}S_{0}$ can be written
as
\begin{eqnarray}
X_{t}^{\psi,\varepsilon} & = & \psi_{t}^{0}+\psi_{t}\mathbf{1}_{\{ \psi_{t}\geq0\} }
(1-\varepsilon)S_{t}+\psi\mathbf{1}_{\{ \psi_{t}<0\} }(1+\varepsilon)S_{t}\nonumber \\
 & = & \psi_{0}^{0}+(1-\varepsilon)S\mal\psi_{t}^{\downarrow}
 -(1+\varepsilon)S\mal\psi_{t}^{\uparrow}\nonumber \\
 &  & +\psi_t\mathbf{1}_{\{ \psi_t\geq0\} }(1-\varepsilon)S_t
 +\psi_t\mathbf{1}_{\{ \psi_t<0\} }(1+\varepsilon)S_t.\label{eq:self-fin.}
\end{eqnarray}
If $(1-\varepsilon)S\mal\psi^{\downarrow}=\widetilde{S}\mal\psi^{\downarrow}$
and $(1+\varepsilon)S\mal\psi^{\uparrow}=\widetilde{S}\mal\psi^{\uparrow}$
for some Itô process $\widetilde{S}$ with values in $[(1-\varepsilon)S,(1+\varepsilon)S]$,
then (\ref{eq:self-fin.}) and integration by parts yield
\begin{eqnarray*}
X_{t}^{\psi,\varepsilon} & = & \psi_{0}^{0}-\widetilde{S}\mal\psi_{t}
+\psi_{t}\widetilde{S}_{t}-\psi_{t}\mathbf{1}_{\{ \psi_{t}\geq0\} }(\widetilde{S}_{t}
-(1-\varepsilon)S_{t})-\psi_{t}\mathbf{1}_{\{ \psi_{t}<0\} }
(\widetilde{S}_{t}-(1+\varepsilon)S_{t})\\
 & = & \psi_{0}^{0}+\psi_{0}\widetilde{S}_{0}+\psi\mal\widetilde{S}_{t}
 -\psi_{t}\mathbf{1}_{\{ \psi_{t}\geq0\} }(\widetilde{S}_{t}
 -(1-\varepsilon)S_{t})-\psi_{t}\mathbf{1}_{\{ \psi_{t}<0\} }
 (\widetilde{S}_{t}-(1+\varepsilon)S_{t})
\end{eqnarray*}
and hence
\[
\big|X_{t}^{\psi,\varepsilon}-(x+\psi\mal\widetilde{S}_{t})\big|\leq\varepsilon x^{S}+2\varepsilon|\psi_{t}S_{t}|.
\]
\end{remark}

In this setting, we focus on the exponential utility function with
constant absolute risk aversion $p>0$:
\[
\uti(x):=-e^{-px}.
\]
Our optimization problem consists in maximizing the expected utility
or, equivalently, the certainty equivalent $\mathbf{CE}(X_{T}^{\psi,\varepsilon})$
of terminal wealth over all admissible trading strategies $\psi$
with given initial wealth $(x^{B},x^{S})$. As usual, the\emph{
certainty equivalent} of a random payoff $X$ refers to the deterministic
amount with the same utility, i.e.,
\[
\mathbf{CE}(X):=-\frac{1}{p}\ln\mathbf{E}\big[e^{-pX}\big].
\]

\section{Main results} \label{sec:MainResults}
In this section, we present the main theorem of this paper concerning
optimal investment to the leading order. To this end, we require that
the corresponding frictionless market fulfills some regularity conditions.

\begin{assumption} \label{ass:MEMM}
We suppose that the frictionless price process $S$ allows for an
equivalent local martingale measure with finite relative entropy.
\end{assumption}

Denote the initial wealth before liquidation by $x:=x^{B}+x^{S}$.
According to \cite[Theorem 2.1]{frittelli.00}, Assumption~\ref{ass:MEMM} implies that the \emph{minimal
entropy (local) martingale measure} ($MEMM$) $\mathbf{Q}$ for $S$
exists. By \cite[Theorem 2.2 (iv)]{schachermayer.99}, there is a predictable,
$S$-integrable process $\varphi$ such that $\varphi\mal S$
is a $\mathbf{Q}$-martingale and
\[
\frac{d\mathbf{Q}}{d\mathbf{P}}=\frac{\uti^{\prime}(x+\varphi\mal S_{T})}{y}
\]
with $y:=\mathbf{E}[\uti^{\prime}(x+\varphi\mal S_{T})]$.
Interpreted as number of shares, strategy $\varphi$ is the optimal
solution to the frictionless counterpart of the above utility maximization
problem.

For any Itô process $X$, we denote by $b^{X}$ and $c^{X,X}$ its
local $\mathbf{Q}$-drift and quadratic variation, respectively, i.e.,
\[
dX_{t}=b_{t}^{X}dt+dM_{t}^{X,\mathbf{Q}},\qquad c_{t}^{X,X}:=\frac{d[X,X]_{t}}{dt},
\]
where $M^{X,\mathbf{Q}}$ is a continuous $\mathbf{Q}$-local martingale
starting in $0$. Similarly, for Itô processes $X$ and $Y$, their
local covariation is denoted by
\[
c_{t}^{X,Y}:=\frac{d[X,Y]_{t}}{dt}.
\]
The local drift rate of Itô process $X$ relative to $\mathbf{P}$ rather
than $\mathbf{Q}$ is written as $b^{X,\mathbf{P}}$.

\begin{assumption} \label{ass:ItoDynamic}
We suppose that the frictionless optimizer $\varphi$ and the \emph{activity
rate}
\[
\varrho:=\frac{c^{\varphi,\varphi}}{c^{S,S}}
\]
are well-defined Itô processes such that $\varrho$ never vanishes.
\end{assumption}

The processes
$S$, $\varphi$, $\varrho$ and their dynamics depend on the current
level of the stock price. In concrete models, the following related
normalized processes are easier to work with:
\begin{itemize}
\item the \emph{stock return process} $R:=\ln S$,
\item the \emph{stock holdings} $\pi:=\varphi S$,
\item the \emph{normalized activity rate} $\eta:=\varrho S^{4}$.
\end{itemize}

\begin{assumption}\label{ass:all}
We assume that
\begin{equation}
\mathbf{E_{Q}}\left[\sup_{t\in[0,T]}|X_{t}|^{n}\right]<\infty
\quad\textrm{for any }n\in\mathbb{N}\textrm{ and  any }X\in\mathcal{H},\label{eq:Bed.-all}
\end{equation}
where
\[
\mathcal{H}:=\{ \pi,\eta,\eta^{-1},b^\pi,b^{\eta},c^{R,R},(c^{R,R})^{-1},c^{\pi,\pi},c^{\eta,\eta}\} .
\]
Moreover, we suppose
\begin{equation}
\mathbf{E_{Q}}[\exp(|9p\varphi\mal S_{T}|)]<\infty.\label{eq:Bed.U_2}
\end{equation}
Finally, we assume that $c^{\pi,\pi},c^{R,R},c^{\eta,\eta},c^{\pi,\eta},c^{\pi,R},c^{\eta,R}$ are continuous.
\end{assumption}

\begin{thm}\label{thm:MainTheorem}
Suppose that Assumptions~\ref{ass:MEMM}, \ref{ass:ItoDynamic}, \ref{ass:all} hold.  Let
\[
\Delta\varphi^{\pm}:=\pm\left(\frac{3\varrho}{2p}S\varepsilon\right)^{1/3}
=\pm\left(\frac{3\eta}{2p}\varepsilon\right)^{1/3}S^ {-1}.
\]
\begin{enumerate}
\item 
There exists a continuous adapted process
\[
\varphi^{\varepsilon}=\varphi+\Delta\varphi=\varphi^{\varepsilon\uparrow}-\varphi^{\varepsilon\downarrow},
\]
 where $\Delta\varphi$ has values in $[\Delta\varphi^{-},\Delta\varphi^{+}]$,
\[
\varphi_{0}^{\varepsilon}=\begin{cases}
\varphi_{0}+\Delta\varphi_{0}^{+} & \textrm{if }x^{S}>(\varphi_{0}+\Delta\varphi_{0}^{+})S_{0,}\\
\varphi_{0}+\Delta\varphi_{0}^{-} & \textrm{if }x^{S}<(\varphi_{0}+\Delta\varphi_{0}^{-})S_{0},\\
{x^{S}/S_{0}} & \textrm{otherwise},
\end{cases}
\]
and $\varphi^{\varepsilon\uparrow}$, $\varphi^{\varepsilon\downarrow}$
are increasing process such that
\[
\varphi^{\varepsilon\uparrow}\:\textrm{increases only on the set}\:\{ \Delta\varphi=\Delta\varphi^{-}\} 
\subseteq\Omega\times[0,T],
\]
\[
\varphi^{\varepsilon\downarrow}\:\textrm{increases only on the set}\:\{ \Delta\varphi=\Delta\varphi^{+}\} 
\subseteq\Omega\times[0,T].
\]

	\item 
	By slight abuse of notation, we identify $\varphi^{\varepsilon}$ with the unique self-financing strategy
	$(\psi^0,\psi)$ that satisfies
	$\psi^0_0=x^B$, $\psi_0S_0=x^S$, $\psi_t=\varphi^{\varepsilon}_t$
	for $t\in(0,T]$.
	Define
\begin{equation}\label{e:tauepsilon}
 \tau^{\varepsilon}:=\inf\left\{ t\in[0,T]:|X_{t}^{\varphi^{\varepsilon},\varepsilon}
-(x+\varphi\mal S_{t})|>1\mbox{ or }|X_{t}^{\varphi^{\varepsilon},\varepsilon}|
>\varepsilon^{-4/3}\right\} \wedge T.
\end{equation}
Then $\mathbf{P}(\tau^{\varepsilon}=T)\to1$
as $\varepsilon\to0$. Moreover, $\varphi^{\varepsilon}\mathbf{1}_{\auf 0,\tau^{\varepsilon}\zu}$
is a utility-maximizing strategy to the leading order $O(\varepsilon^{2/3})$,
i.e.,
\[
\sup_{\psi\in\mathcal{A}^{\varepsilon}(x^{B},x^{S})}
\mathbf{E}\left[\uti(X_{T}^{\psi,\varepsilon})\right]
=\mathbf{E}\left[\uti(X_{\tau^{\varepsilon}}^{\varphi^{\varepsilon},\varepsilon})\right]
+o(\varepsilon^{2/3}).
\]
(As above, $\varphi^{\varepsilon}\mathbf{1}_{\auf 0,\tau^{\varepsilon}\zu}$ here refers to the strategy 
$\psi\in\mathcal{A}^{\varepsilon}(x^{B},x^{S})$ with 
$\psi_t=\varphi^{\varepsilon}_t\mathbf{1}_{\auf 0,\tau^{\varepsilon}\zu}(t)$
for $t\in(0,T]$.)
	\item The optimal certainty equivalent amounts to
\begin{eqnarray*}
 \sup_{\psi\in\mathcal{A}^{\varepsilon}(x^{B},x^{S})}\mathbf{CE}(X_{T}^{\psi,\varepsilon})
&=&\mathbf{CE}(X_{\tau^{\varepsilon}}^{\varphi^{\varepsilon},\varepsilon})+o(\varepsilon^{2/3})\\
&=&\mathbf{CE}(x+\varphi\mal S_{T})-\frac{p}{2}
\mathbf{E_{Q}}\left[(\Delta\varphi^{+})^{2}\mal[S,S]_{T}\right]+o(\varepsilon^{2/3}).
\end{eqnarray*}
\end{enumerate} \end{thm}

\begin{prf}
The proof is split up into several steps given in Section~\ref{sec:Proof}. The existence
of $\varphi^{\varepsilon}$ is linked to the Skorohod problem with
time-dependent reflecting barriers (cf.\ Lemma~\ref{lem:Refl.-Prinzip}). With the help of
the shadow price process $S^{\varepsilon}$ derived heuristically
in \cite{kallsen.muhlekarbe.12} (cf.\ Corollary~\ref{lem:Ex.ShadowPrice}), the utility generated by $\varphi^{\varepsilon}$
stopped at $\tau^{\varepsilon}$ is computed in Lemma~\ref{lem:Primal}. The optimality
of $\varphi^{\varepsilon}$ is proved by means of some dual considerations
(cf.\ Lemma~\ref{lem:Dual}) in conjunction with the conjugate relation (cf.\ Lemma~\ref{thm:Opt}). Finally, the proof of the explicit expression for the certainty
equivalent loss relies on a random time change and the ergodic property
of reflected Brownian motion (cf.\ Corollary~\ref{cor:Nutzenverlust}).
\end{prf}

\begin{remark}\label{rem:Bed}
Roughly speaking, the assumptions in Theorem \ref{thm:MainTheorem} 
concern sufficient integrability of the solution
to the frictionless utility maximization problem in order to warrant
that the maximal expected utility is twice differentiable as a function
of $\varepsilon^{1/3}$. In the subsequent section we verify
these assumptions in a general stochastic volatility setup.
\end{remark}

From our theorem, the leading-order optimal strategy under transaction costs $\varphi^{\varepsilon}$
stays within the random no-trade region $[\varphi+\Delta\varphi^-,\varphi+\Delta\varphi^+]$ around
the frictionless optimizer $\varphi$; and it increases (resp.\,decreases)
only while hitting the lower (resp.\,upper) bound. In this sense,
$\varphi+\Delta\varphi^{+}$ and $\varphi+\Delta\varphi^{-}$ correspond to the \emph{selling}
and \emph{buying boundary}, respectively. At the random
time $\tau^{\varepsilon}$, the portfolio is liquidated primarily in order to
bound losses.

\section{Examples}\label{sec:Examples}
We provide two classes of models for which the frictionless optimizer
$\varphi$ is known explicitly.

\subsection{Black-Scholes model}
First, we consider the so-called \emph{Black-Scholes model}
\[
dS_{t}=S_{t}(bdt+\sigma dW_{t})
\]
with $b\in\mathbb R, \sigma\in\mathbb{R}_{+}\setminus\{ 0\}$.
We show that Assumptions \ref{ass:MEMM}, \ref{ass:ItoDynamic},
\ref{ass:all} hold if $b\neq0$.

From Theorem~\ref{thm:Volphi} in the appendix, the frictionless optimal strategy
$\varphi$ satisfies
\[
\pi_{t}=\varphi_{t}S_{t}=\frac{b}{p\sigma^{2}}\quad\textrm{for all }t\in[0,T].
\]
By Itô's formula, we have
\[
d\varphi_{t}=-\frac{b}{p\sigma^{2}S_{t}^{2}}dS_{t}+\frac{b}{p\sigma^{2}S_{t}^{3}}d[S,S]_{t}
\]
and hence
\[
c_{t}^{\varphi,\varphi}=\frac{b^{2}}{p^{2}\sigma^{2}S_{t}^{2}}.
\]
This yields
\[
\varrho_{t}=\frac{b^{2}}{p^{2}\sigma^{4}S_{t}^{4}}
\]
and 
\[
\eta_{t}=\frac{b^{2}}{p^{2}\sigma^{4}}
\]
for any $t\in[0,T]$.
Therefore, all processes in set $\mathcal{H}$ as well as
$c^{R,\pi},c^{R,\eta},c^{\pi,\eta}$
are constant, which in particular
yields Condition (\ref{eq:Bed.-all}). The frictionless
optimal terminal gains are of the form
\[
\varphi\mal S_{T}=\frac{b}{p\sigma}W_{T}^{\mathbf{Q}},
\]
where $W_{t}^{\mathbf{Q}}=W_{t}+\frac{b}{\sigma}t$, $t\in[0,T]$
is a standard Brownian motion under measure $\mathbf{Q}$.
Thus Condition
(\ref{eq:Bed.U_2}) is satisfied. 

The no-trade bounds are obtained from
\begin{equation}\label{e:bsnt}
 \Delta\varphi^{\pm}S=\pm\left(\frac{3b^{2}\varepsilon}{2p^{3}\sigma^{4}}\right)^{1/3}
\end{equation}
and the certainty equivalent loss due to transaction costs is
\begin{equation}\label{e:bsloss}
\sup_{\psi\in\mathcal{A}^{\varepsilon}(x^{B,}x^{S})}
\mathbf{CE}(X_{T}^{\psi,\varepsilon})-\mathbf{CE}(x+\varphi\mal S_{T})
=-\left(\frac{9b^{4}\varepsilon^{2}}{32\sigma^2}\right)^{1/3}\frac Tp+o(\varepsilon^{2/3}).
\end{equation}
(\ref{e:bsnt}) coincides with the formulas in \cite[p.319]{whalley.willmott.97} resp.\
\cite[(3.4)]{bichuch.11b}. 
The expression in (\ref{e:bsloss}), on the other hand,
is obtained from \cite[(3.7, 3.8)]{bichuch.11b} if one uses the equation for $V_2$ in
\cite[p.317]{whalley.willmott.97}.
Note, however, that \cite{bichuch.11b} considers a slightly more involved notion
of admissibility.

\subsection{Stochastic volatility model}\label{subsec:Vol-Modell}

Let us turn to the following stochastic volatility model:
\begin{equation}
dS_{t}=S_{t}(b(Y_{t})dt+\sigma(Y_{t})dW_{t})\label{eq:verallg. BS}
\end{equation}
with continuous functions $b,\sigma:\mathbb{R}\to\mathbb{R}$
and an Itô process $Y$ which is independent of the Brownian motion $W$. The filtration
is supposed to be generated by $W$ and $Y$.
\begin{prop}
Suppose that the stochastic volatility model (\ref{eq:verallg. BS}) is such that
\begin{itemize}
\item 
the coefficients $b(Y)$, $\sigma(Y)$ are bounded
processes that are  bounded away from $0$,
\item the processes
\begin{equation}\label{e:pieta}
\pi:=\frac{b(Y)}{p\sigma(Y)^{2}},\qquad
\eta:=\pi^{2}+\frac{c^{\pi,\pi}}{\sigma(Y)^{2}},
\end{equation}
are Itô processes with bounded coefficients 
$b^{\pi,\mathbf{P}},c^{\pi,\pi},  b^{\eta,\mathbf{P}}, c^{\eta,\eta}$
and continuous coefficients $c^{\pi,\pi},  c^{\eta,\eta}, c^{\pi,\eta},$
\item for
\begin{equation}
\widetilde{Z}_{t}:=\mathbf{E}\!\left[\left.\exp\left(-\frac{1}{2}\int_{0}^{T}\left(\frac{b(Y_{t})}
{\sigma(Y_{t})}\right)^{2}dt\right)\right|\mathcal{F}_{t}\right],\label{eq:verallg. BS-Z}
\end{equation}
the process ${c^{\widetilde{Z},\widetilde{Z}}}/{\widetilde{Z}^{2}}$ is bounded.
\end{itemize}
Then Assumptions \ref{ass:MEMM}, \ref{ass:ItoDynamic}, \ref{ass:all} are satisfied. Moreover, 
$\pi$ and $\eta$ are the corresponding stock holdings and normalized activity rate.
The no-trade boundary is given by
\[
\Delta\varphi^{\pm}S=\pm\left(\frac{3\eta\varepsilon}{2p}\right)^{1/3}
\]
and the certainty equivalent loss amounts to
\[
\mathbf{E_{Q}}\left[\int_{0}^{T}\left(\frac{9p\eta_t^{2}}{32}\right)^{1/3}
\sigma(Y_{t})^{2}dt\right]\varepsilon^{2/3}+o(\varepsilon^{2/3}).
\]
\end{prop}

\begin{prf}
\emph{Step 1:}
We show that $[f(S),X]=0$
for any $C^{2}$-function $f$ and any Itô process $X$ which
is $\sigma(Y)$-measurable.
Indeed, by Itô's formula it suffices to prove that $[W,X]=0$.
Using \cite[Theorem II.4]{protter.04}, it is easy to show that the
martingale part of $X$ is $\sigma(Y)$-measurable. Hence
without loss of generality, $X$ is a local martingale. By localization
it suffices to consider the case where $X$ is a square-integrable
martingale. 
Let $\mathbf{G}$ be the filtration defined
in (\ref{eq:Filt.G}) in the appendix.
Then $X_{t}$ is $\mathcal{G}_{0}$-measurable
for any $t\in[0,T]$ and $W$ is a Brownian motion relative
to both $\mathbf{F}$ and $\mathbf{G}$. We obtain 
\begin{eqnarray*}
\mathbf{E}[W_{t}X_{t}|\mathcal{F}_{s}] 
& = & \mathbf{E}[\mathbf{E}[W_{t}X_{t}|\mathcal{G}_{s}]|\mathcal{F}_{s}]\\
& = & \mathbf{E}[\mathbf{E}[W_{t}|\mathcal{G}_{s}]X_{t}|\mathcal{F}_{s}]\\
& = & W_{s}\mathbf{E}[X_{t}|\mathcal{F}_{s}]\\
& = & W_{s}X_{s}
\end{eqnarray*}
for any $s<t$.
Hence $WX$ is a martingale, which implies
$[W,X]=0$ as desired.

\emph{Step 2:}
We show that $\pi,\eta$ in (\ref{e:pieta}) coincide with the stock holdings and the normalized activity rate.
By Theorem~\ref{thm:Volphi}, the frictionless optimizer $\varphi$ satisfies
$\varphi_{t}S_{t}=\pi_{t}$
for any $t\in[0,T]$.
From Itô's formula we get
\[
d\varphi_{t}=-\frac{\pi_{t}}{S_{t}^{2}}dS_{t}
+\frac{\pi_{t}}{S_{t}^{3}}d[S,S]_{t}+\frac{1}{S_{t}}d\pi_{t}+d\left[\frac{1}{S},\pi\right]_{t}.
\]
Step 1 yields $[S,\pi]=0$, which implies
\[
c_{t}^{\varphi,\varphi}=\frac{\pi_{t}^{2}\sigma(Y_{t})^{2}+c_{t}^{\pi,\pi}}{S_{t}^{2}}
\]
and
\[
\frac{c_{t}^{\varphi,\varphi}}{c^{S,S}_t}S_{t}^{4}
=\pi_{t}^{2}+\frac{c_{t}^{\pi,\pi}}{\sigma(Y_{t})^{2}}
=\eta_{t}.
\]

\emph{Step 3:}
Let $\overline Z$ be defined as in (\ref{e:Zquer}).
By Theorem \ref{thm:Volphi},
$\widetilde{Z}\overline Z/\widetilde{Z}_0$ is the density process 
of the  MEMM $\mathbf Q$.
For any It\^o process $X$, Girsanov's theorem implies 
\[
b_{t}^{X}=b_{t}^{X,\mathbf{P}}
+\frac{c_{t}^{\widetilde{Z}\overline{Z},X}}{\widetilde{Z}_t\overline{Z}_{t}}.
\] 
Integration by parts yields 
\begin{eqnarray*}
\left|\frac{c^{\widetilde{Z}\overline{Z},X}}{\widetilde{Z}\overline{Z}} \right|
&=&\left|\frac{c^{\widetilde{Z},X}}{\widetilde{Z}} 
+\frac{c^{\overline{Z},X}}{\overline{Z}} \right|\\
&\leq&\sqrt{\frac{c^{\widetilde{Z},\widetilde{Z}}}{\widetilde{Z}^2}}\sqrt{c^{X,X}}
+\sqrt{\frac{c^{\overline{Z},\overline{Z}}}{\overline{Z}^2}}\sqrt{c^{X,X}}\\
&=&\left(\sqrt{\frac{c^{\widetilde{Z},\widetilde{Z}}}{\widetilde{Z}^2}}
+\left|\frac{b(Y)}{\sigma(Y)}\right|\right)\sqrt{c^{X,X}}.
\end{eqnarray*}
In view of our boundedness assumptions, we conclude that $b^\pi$ and $b^\eta$ are bounded.
Consequently, all processes in set $\mathcal{H}$ are
bounded, which implies Condition (\ref{eq:Bed.-all}).

Moreover, the
frictionless optimal terminal gains are of the form
\[
\varphi\mal S_{T}=\int_{0}^{T}\frac{b(Y_{t})}{p\sigma(Y_{t})}dW_{t}^{\mathbf{Q}},
\]
where $W^{\mathbf{Q}}=W+\int_{0}^{\cdot}\frac{b(Y_{t})}{\sigma(Y_{t})}dt$
is a standard Brownian motion under measure $\mathbf{Q}$.
If an integrand $H$ is bounded by $m\in\mathbb R$, we have
\begin{eqnarray*}
 \mathbf{E}_\mathbf{Q}\left[\exp(H\mal W^\mathbf{Q}_T)\right]
 &\leq&\mathbf{E}_\mathbf{Q}\left[\exp\bigg(H\mal W^\mathbf{Q}_T-\frac12\int_0^TH_t^2dt\bigg)\right]
\exp\left(\frac12m^2T\right)\\
&\leq& \exp\left(\frac12m^2T\right) <\infty.
\end{eqnarray*}
Together, we conclude that Condition (\ref{eq:Bed.U_2}) holds.
Finally, Step 1 yields that $c^{\pi,R}=0$ and  $c^{\eta,R}=0$, 
which completes the proof of Assumption \ref{ass:all}.
\end{prf}

\section{Proof of the main results}\label{sec:Proof}
As indicated in Section~\ref{sec:MainResults}, we prove the main theorem in this section. 
We assume throughout that Assumptions~\ref{ass:MEMM}, \ref{ass:ItoDynamic} hold. 
The idea of our proof can be outlined as follows:

\begin{itemize}
\item \cite{kallsen.muhlekarbe.12} derives a possibly suboptimal candidate strategy
$\varphi^{\varepsilon}$ (in the sense of number of shares of stock)
along with a \emph{shadow price} $S^{\varepsilon}$. This term here
refers to a frictionless price process moving within the bid-ask bounds
$[(1-\varepsilon)S,(1+\varepsilon)S]$
and such that strategy $\varphi_{t}^{\varepsilon}$ only buys (resp.\,sells)
stock if the shadow price $S_{t}^{\varepsilon}$ coincides with the
ask price $(1+\varepsilon)S_{t}$ (resp.\,bid price $(1-\varepsilon)S_{t}$).
Evidently, following $\varphi^{\varepsilon}$ in the frictionless
market $S^{\varepsilon}$ yields the same wealth process and hence
expected utility as in the original market with proportional transaction
costs. This expected utility can be computed explicitly to the leading
order because both $\varphi^{\varepsilon}$ and $S^{\varepsilon}$
are given in closed form.
\end{itemize}

\begin{itemize}
\item According to \cite{karatzas.al.91,schachermayer.99} dealing with the issue of hedging
duality in frictionless markets, the utility maximization problem
for $S^{\varepsilon}$ without transaction costs is related to a dual
minimization problem on the set of equivalent local martingale measures.
Specifically, the value of the dual problem dominates the expected
utility of any admissible trading strategy. In a second step we therefore
construct a carefully chosen, explicitly known local martingale measure
(identified with its Radon-Nikodym density $Z^{\varepsilon}$) for
$S^{\varepsilon}$. Since trading in the frictionless market $S^{\varepsilon}$
leads to higher profit than in the original market with transaction
costs, the Lagrange dual function evaluated at $Z^{\varepsilon}$
provides an upper bound to the maximal expected utility in the 
market with friction. This upper bound can be computed explicitly to the leading
order because $Z^{\varepsilon}$ is known in closed form.
\end{itemize}

\begin{itemize}
\item In a final step we observe that the suboptimal expected utility of
$\varphi^{\varepsilon}$ coincides to the leading order with the upper
bound above. Hence, we obtain approximate optimality of the candidate
strategy.
\end{itemize}

In the language of \cite{cvitanic.karatzas.96}, $(Z^{\varepsilon},Z^{\varepsilon}S^{\varepsilon})$
is a \emph{state-price density}, which, by duality to the set of self-financing
portfolios in the market with friction, provides an upper bound to
the expected utility under transaction costs.

Set
\[
\alpha:=\frac{pc^{S,S}}{3c^{\varphi,\varphi}},\qquad
\beta:=\left(\frac{S}{\alpha}\right)^{1/3}
=\Delta\varphi^+\left(\frac{2}{\varepsilon}\right)^{1/3}.
\]
We define sets of processes
\[
\mathcal{H}^{b^{\Delta S}}
:=\left\{ \beta c^{S,S},\alpha\beta^{2}b^{\varphi},
\beta^{2}c^{\alpha,\varphi},\beta^{3}b^{\alpha},\beta b^{\alpha\beta^{2}},c^{\alpha\beta^{2},\varphi}\right\} ,
\]
\[
\mathcal{H}^{c^{\Delta S,S}}
:=\left\{ \alpha\beta^{2}c^{\varphi,S},\beta^{3}c^{\alpha,S},\beta c^{\alpha\beta^{2},S}\right\} ,
\]
\[
\mathcal{H}^{c^{\Delta S,\Delta S}}
:=\left\{ \alpha^{2}\beta^{4}c^{\varphi,\varphi},\beta^{6}
c^{\alpha,\alpha},\beta^{2}c^{\alpha\beta^{2},\alpha\beta^{2}}\right\} ,
\]
\[
\mathcal{H}^{c^{\Delta S,\varphi}}:=\left\{ \alpha\beta^{2}c^{\varphi,\varphi},
\beta^{3}c^{\alpha,\varphi},\beta c^{\alpha\beta^{2},\varphi}\right\} ,
\]
\[
\mathcal{G}_{1}:=\{ Sb^{\varphi}\} \cup\mathcal{H}^{c^{\Delta S,\varphi}}\cup
\left\{ \beta b^{\Delta}:b^{\Delta}\in\mathcal{H}^{b^{\Delta S}}\right\} ,
\]
\[
\mathcal{G}_{2}:=\left\{ \beta^{2}c^{S,S},S^{2}c^{\varphi,\varphi}\right\} \cup
\left\{ \beta^{2}c^{\Delta,\Delta}:c^{\Delta,\Delta}\in\mathcal{H}^{c^{\Delta S,\Delta S}}\right\} ,
\]
\[
\mathcal{G}:=\mathcal{G}_{1}\cup\mathcal{G}_{2}.
\]

For the proof of Theorem \ref{thm:MainTheorem}, Assumption \ref{ass:all} can be replaced by the 
following two slightly weaker assumptions.
\begin{assumption}\label{ass:Primal}
We suppose that
\begin{equation}
\sum_{g\in\mathcal{G}}\mathbf{E_{Q}}\left[\int_{0}^{T}|g_{t}|^{4}dt\right]
+\mathbf{E_{Q}}\left[\sup_{t\in[0,T]}|\varphi_{t}S_{t}|^{4}+\sup_{t\in[0,T]}|\beta_{t}S_{t}|^{4}\right]
<\infty,\label{eq:Bed.Primal}
\end{equation}
\begin{equation}
\sum_{c^{\Delta,S}\in\mathcal{H}^{c^{\Delta S,S}}}
\mathbf{E_{Q}}\left[\sup_{t\in[0,T]}\left|\frac{c_{t}^{\Delta,S}}{c_{t}^{S,S}} \right|^{16}\right]
<\infty,\label{eq:Bed.Dual_sigma}
\end{equation}
\begin{equation}
\sum_{b^{\Delta}\in\mathcal{H}^{b^{\Delta S}}}
\mathbf{E_{Q}}\left[\int_{0}^{T}\left|\frac{b_{t}^{\Delta}}{\sigma_{t}^{S}}\right|^{16}dt\right]
<\infty,\label{eq:Bed.Dual}
\end{equation}
\[
\mathbf{E_{Q}}[\exp(|9p\varphi\mal S_{T}|)]<\infty.
\]
\end{assumption}

\begin{assumption}\label{ass:Cont.}
The processes $c^{\varphi,\varphi},c^{S,S}, c^{\beta,\beta},c^{\varphi,\beta}$ are continuous
and hence pathwise bounded. Moreover,
the processes $b^\varphi, b^\beta$ are assumed to be pathwise bounded as well.
\end{assumption}

\begin{lemma}\label{lem:Alg.}
Define
\[
\mathcal{X}:=\left\{ (X_{t})_{t\in[0,T]}:\mathbf{E_{Q}}\bigg[\sup_{t\in[0,T]}|X_{t}|^{n}\bigg]
<\infty\textrm{ for any }n\in\mathbb{N}\right\} .
\]
Then
\begin{enumerate}
\item For $X,Y\in\mathcal{X}$, $c\in\mathbb{R}$, it holds that $X+Y,XY,cX\in\mathcal{X}$.
\item If $X\in\mathcal{X}$ and $f:\mathbb{R}\to\mathbb{R}$
with $|f(x)|\leq1+|x|$ for any $x\in\mathbb{R}$,
then $f(X)\in\mathcal{X}$.
\item If $X\in\mathcal{X}$, then $\mathbf{E_{Q}}[\int_{0}^{T}|X_{t}|^{n}dt]<\infty$
for any $n\in\mathbb{N}$.
\end{enumerate}
\end{lemma}

\begin{prf}
This is straightforward.
\end{prf}

\begin{lemma}
Assumptions~\ref{ass:Primal} and \ref{ass:Cont.} hold if Assumptions \ref{ass:MEMM}-\ref{ass:all} are fulfilled.
\end{lemma}

\begin{prf}
This follows from It\^o's  formula,
straightforward but tedious calculations, and Lemma \ref{lem:Alg.}.
\end{prf}

\subsection{Existence of shadow price $S^\varepsilon$}
\begin{lemma}\label{lem:Refl.-Prinzip}
For
\[
\Delta\varphi_{0}:=\begin{cases}
\Delta\varphi_{0}^{+} & \textrm{if }x^{S}>(\varphi_{0}+\Delta\varphi_{0}^{+})S_{0},\\
\Delta\varphi_{0}^{-} & \textrm{if }x^{S}<(\varphi_{0}+\Delta\varphi_{0}^{-})S_{0},\\
\frac{x^{S}}{S_{0}}-\varphi_{0} & \textrm{otherwise},
\end{cases}
\]
there exists a solution $\Delta\varphi$
to the Skorohod stochastic differential equation (SDE)
\begin{equation}
d\Delta\varphi_{t}=-d\varphi_{t}\label{eq:Dynamik_phi}
\end{equation} with reflection at $\Delta\varphi^{-}$, $\Delta\varphi^{+}$,
i.e., there exist a continuous, adapted, $[\Delta\varphi^{-},\Delta\varphi^{+}]$-valued
process $\Delta\varphi$ and adapted increasing processes $\varphi^{\varepsilon\uparrow}$,
$\varphi^{\varepsilon\downarrow}$ such that
\begin{equation}
\varphi^{\varepsilon\uparrow}\textrm{ increases only on the set }
\{ \Delta\varphi=\Delta\varphi^{-}\} \subseteq\Omega\times[0,T],\label{eq:Rand_phi_+}
\end{equation}
\begin{equation}
\varphi^{\varepsilon\downarrow}\textrm{ increases only on the set }
\{ \Delta\varphi=\Delta\varphi^{+}\} \subseteq\Omega\times[0,T],\label{eq:Rand_phi_-}
\end{equation}
and
\begin{equation}
\Delta\varphi=-\varphi+\varphi^{\varepsilon\uparrow}-\varphi^{\varepsilon\downarrow}.\label{eq:Def.delta-phi}
\end{equation}
\end{lemma}

\begin{prf}
(\ref{eq:Dynamik_phi}) is a SDE related to the semimartingale
$\varphi$ with constant coefficients which are obviously Lipschitz continuous
with respect to $\Delta\varphi$. The time-dependent reflecting barriers
$\Delta\varphi^{\pm}$ are Lipschitz operators in the sense of \cite[Definition 3.1]{slominski.wojciechowski.13}
evaluated at process $\Delta\varphi$. The
assertion follows now from \cite[Theorem 3.3]{slominski.wojciechowski.13}.
\end{prf}

\begin{cor}\label{lem:Ex.ShadowPrice}
Let $\Delta\varphi,\varphi^{\varepsilon\uparrow},\varphi^{\varepsilon\downarrow}$
be as in Lemma~\ref{lem:Refl.-Prinzip} resp.\ Theorem~\ref{thm:MainTheorem}. Define
 \[\gamma:=3\alpha\beta^{2}\left(\frac\varepsilon2\right)^{2/3}\]
 and
\begin{equation}
\Delta S:=\alpha\Delta\varphi^{3}-\gamma\Delta\varphi.\label{eq:Def.Delta-S}
\end{equation}
Then $\Delta S$ is an Itô process with values in $[-\varepsilon S,\varepsilon S]$
such that
\begin{equation}
\varphi^{\varepsilon\uparrow}\textrm{ increases only on the set }
\{ \Delta S=\varepsilon S\} \subseteq\Omega\times[0,T],\label{eq:Rand_delta-S_+}
\end{equation}
\begin{equation}
\varphi^{\varepsilon\downarrow}\textrm{ increases only on the set }
\{ \Delta S=-\varepsilon S\} \subseteq\Omega\times[0,T].\label{eq:Rand_delta-S_-}
\end{equation}
\end{cor}

\begin{prf}
Consider the function 
$f(x,a,g):=a x^{3}-g x$.
By Assumption \ref{ass:ItoDynamic}, $\alpha,\gamma,\varphi$ are Itô processes.
From (\ref{eq:Rand_phi_+}, \ref{eq:Rand_phi_-}, \ref{eq:Def.delta-phi})
we deduce that
\begin{equation}
d\Delta\varphi_{t}=-d\varphi_{t}\textrm{ on }\{ \Delta\varphi\neq\Delta\varphi^{\pm}\}.\label{eq:Dynamik_Delta-phi}
\end{equation}
Since
\begin{equation}
\frac{\partial f}{\partial x}(\Delta\varphi^{\pm},\alpha,\gamma)=3\alpha(\Delta\varphi^{\pm})^{2}-\gamma=0,
\label{eq:Rand_phi-gamma}
\end{equation}
Itô's formula yields 
\begin{eqnarray*}
d\Delta S_{t} &=& df(\Delta\varphi,\alpha,\gamma)_t\\
& = & -(3\alpha_{t}\Delta\varphi_{t}^{2}
-\gamma_{t})d\varphi_{t}+3\alpha_{t}\Delta\varphi_{t}d[\varphi,\varphi]_{t}\\
 &  & -3\Delta\varphi_{t}^{2}d[\alpha,\varphi]_{t}+\Delta\varphi_{t}^{3}d\alpha_{t}
 -\Delta\varphi_{t}d\gamma_{t}+d[\gamma,\varphi]_{t}.
\end{eqnarray*}
In particular,
$\Delta S$
is an Itô process.
Moreover, (\ref{eq:Rand_phi_+}, \ref{eq:Rand_phi_-}, \ref{eq:Def.Delta-S}) 
and $f(\Delta\varphi^{\pm},\alpha,\gamma)=\mp\varepsilon S$
imply (\ref{eq:Rand_delta-S_+}, \ref{eq:Rand_delta-S_-}).
\end{prf}

The coefficients related to $\Delta S$ can be estimated%
\footnote{Here and in the sequel, inequalities of the form $A\leq \const B$ are to be 
interpreted in the sense that there
exists a constant $k\in\mathbb{R}$ which does not depend on $\varepsilon$ and
such that $A\leq kB$.} 
as follows:
\begin{equation}
b_{t}^{\Delta S}=\underbrace{p\Delta\varphi_{t}c_{t}^{S,S}}
_{=:G^{(1)}_{t}}-\underbrace{(3\alpha_{t}\Delta\varphi_{t}^{2}-\gamma_{t})
b_{t}^{\varphi}-3\Delta\varphi_{t}^{2}c_{t}^{\alpha,\varphi}
+\Delta\varphi_{t}^{3}b_{t}^{\alpha}-\Delta\varphi_{t}
b_{t}^{\gamma}+c_{t}^{\gamma,\varphi}}_{=:G^{(2)}_{t}},\label{eq:b_delta-S}
\end{equation}
\begin{equation}
\big|G^{(1)}_{t}\big|\leq \const\beta_{t}c_{t}^{S,S}\varepsilon^{1/3},
\qquad\big|G^{(2)}_{t}\big|\leq \const\sum_{b^{\Delta}\in\mathcal{H}^{b^{\Delta S}}
\setminus\{ \beta c^{S,S}\} }\big|b_{t}^{\Delta}\big|\varepsilon^{2/3},\label{eq:b_delta-S <=00003D}
\end{equation}
\begin{equation}
\big|c_{t}^{\Delta S,S}\big|
\leq \const\sum_{c^{\Delta,S}\in\mathcal{H}^{c^{\Delta S,S}}}\big|c_{t}^{\Delta,S}\big|
\varepsilon^{2/3},\label{eq:sigma_delta-S<=00003D}
\end{equation}
\begin{equation}
\big|c_{t}^{\Delta S,\Delta S}\big|\leq \const\sum_{c^{\Delta,\Delta}\in
\mathcal{H}^{c^{\Delta S,\Delta S}}}\big|c_{t}^{\Delta,\Delta}\big|\varepsilon^{4/3}\label{eq:c_delta-S}
\end{equation}
and
\begin{equation}
c_{t}^{\Delta S,\varphi}=\underbrace{-pc_{t}^{S,S}(\Delta\varphi_{t}^{2}
-(\Delta\varphi_{t}^{+})^{2})}_{=:G^{(3)}_{t}}
+\underbrace{\Delta\varphi_{t}^{3}c_{t}^{\alpha,\varphi}
-\Delta\varphi_{t}c_{t}^{\gamma,\varphi}}_{=:G^{(4)}_{t}},\label{eq:c_delta-S,phi}
\end{equation}
\begin{equation}
\big|G^{(3)}_{t}\big|\leq \const|\alpha_{t}\beta_{t}^{2}c_{t}^{\varphi,\varphi}
|\varepsilon^{2/3},\qquad\big|G^{(4)}_{t}\big|\leq \const\sum_{c^{\Delta,\varphi}
\in\mathcal{H}^{c^{\Delta S,\varphi}}\setminus\{ \alpha\beta^{2}
c^{\varphi,\varphi}\} }\big|c_{t}^{\Delta,\varphi}\big|\varepsilon.\label{eq:c_delta-S,phi<=00003D}
\end{equation} 
For $\Delta\varphi$ and $\Delta S$ as in Lemma~\ref{lem:Refl.-Prinzip} and
Corollary~\ref{lem:Ex.ShadowPrice} define
\begin{equation}
\varphi^{\varepsilon}:=\varphi+\Delta\varphi,\label{eq:Def.phi}
\end{equation}
\begin{equation}
S^{\varepsilon}:=S+\Delta S.\label{eq:Def.S}
\end{equation}

\begin{remark}
Due to (\ref{eq:Rand_delta-S_+}, \ref{eq:Rand_delta-S_-}), trading
with $\varphi^{\varepsilon}$ at price $S^{\varepsilon}$ without
friction or at the bid/ask prices $S(1\pm\varepsilon)$ generate the same wealth.
$S^{\varepsilon}$ serves as a proxy to the so-called \emph{shadow
price process} which corresponds to the dual optimizer in the market
with transaction costs, cf.\ the process
$\hat RP=\hat Z_1/\hat Z_0$ in \cite[Theorem 6.1]{cvitanic.karatzas.96}.
\end{remark}

\subsection{Primal considerations}\label{subsec:Primal}
With the help of the shadow price process $S^\varepsilon$, we approximate 
the expected utility generated by the candidate strategy $\varphi^{\varepsilon}$.
From \cite[Section V.2]{protter.04} we recall
the $S^{q}$- and $H^{q}$-norms,
$q\in[1,\infty)$, for an Itô process $X$:
\[
\Vert X\Vert _{S^{q}(\mathbf{Q})}:=\left\Vert \sup_{t\in[0,T]}|X_{t}|\right\Vert _{L^{q}(\mathbf{Q})},
\]
\[
\Vert X\Vert _{H^{q}(\mathbf{Q})}:=\left\Vert \int_{0}^{T}|b_{t}^{X}|dt
+\sqrt{\int_{0}^{T}c_{t}^{X,X}dt}\right\Vert _{L^{q}(\mathbf{Q})}.
\]
To be more precise, \cite{protter.04} requires $X_0=0$ in the definition of the $H^{q}$-norm.
\begin{remark}\label{rem:Norm}
Let $q\in[1,\infty)$. The following inequalities will
be useful.
\begin{enumerate}
 \item Due to \cite[Theorem V.2]{protter.04},
\begin{equation}
\Vert X\Vert _{S^{q}(\mathbf{Q})}\leq \const\Vert X\Vert _{H^{q}(\mathbf{Q})}.
\label{eq:S-Norm<=00003DH-Norm}
\end{equation}
holds if $X_0=0$, where the constant does not depend on $X$.
\item By convexity of the mapping $x\mapsto|x|^{q}$ and Jensen's inequality, we have
\begin{equation}
\left|\sum_{n=1}^{N}Y_{n}\right|^{q}
=N^{q}\left|\sum_{n=1}^{N}\frac{Y_{n}}{N}\right|^{q}
\leq N^{q-1}\sum_{n=1}^{N}|Y_{n}|^{q}\label{eq:convex-q}
\end{equation}
 for any $N\in\mathbb{N}$ and any random variables $Y_1,\dots,Y_N$.
In particular, 
\begin{equation}
\left\Vert \sum_{n=1}^{N}X^{(n)}\right\Vert _{q}^{q}\leq \const\sum_{n=1}^{N}
\left\Vert X^{(n)}\right\Vert _{q}^{q}.\label{eq:3-Eks-Ungl.}
\end{equation}
holds for any $N\in\mathbb{N}$,
Itô processes $X^{(1)},\dots,X^{(N)}$, and
$\Vert \cdot\Vert _{q}\in\{ \Vert \cdot\Vert _{S^{q}(\mathbf{Q})},\Vert \cdot\Vert _{H^{q}(\mathbf{Q})}\} $.
\item For any $q\in[1,\infty)$ and any $g\in L^q([0,T])$, H\"older's inequality yields
\begin{equation}\label{eq:Hoelder-ungl.}
 \left(\int_0^T|g(t)|dt\right)^q
 \leq\| g\|^q_{ L^q([0,T])}\|1\|^q_{ L^r([0,T])}
\leq T^{q-1}\int_0^T|g(t)|^qdt
\end{equation}
with $\frac1q+\frac1r=1$.
Moreover, 
\begin{equation}
\left\Vert \sqrt{Y}\right\Vert _{L^{q}(\mathbf{Q})}^{q}
=\Vert Y\Vert _{L^{q/2}(\mathbf{Q})}^{\frac{q}{2}}
\leq\Vert Y\Vert _{L^{q}(\mathbf{Q})}^{\frac{q}{2}}\label{eq:Caucy-Schwarz-Ungl.}
\end{equation}
for any random variable $Y\in L^{q}(\mathbf{Q})$.
\item  
Gathering the above inequalities, we obtain 
\begin{eqnarray}
 \lefteqn{\left\Vert \int_{0}^{T}\bigg(\sum_{m=1}^{M}X_{t}^{(m)}\bigg)dt+\sqrt{\int_{0}^{T}
 \bigg(\sum_{n=1}^{N}Y_{t}^{(n)}\bigg)dt}\right\Vert _{L^{q}(\mathbf{Q})}^{q}}\nonumber \\
 & \overset{(\ref{eq:convex-q})}{\leq} & \const\left(\left\Vert \int_{0}^{T}\bigg(\sum_{m=1}^{M}
 X_{t}^{(m)}\bigg)dt\right\Vert _{L^{q}(\mathbf{Q})}^{q}+\left\Vert \sqrt{\int_{0}^{T}
 \bigg(\sum_{n=1}^{N}Y_{t}^{(n)}\bigg)dt}\right\Vert _{L^{q}(\mathbf{Q})}^{q}\right)\nonumber \\
 & \overset{(\ref{eq:Caucy-Schwarz-Ungl.})}{\leq} & \const\left(\left\Vert \int_{0}^{T}
 \bigg(\sum_{m=1}^{M}X_{t}^{(m)}\bigg)dt\right\Vert _{L^{q}(\mathbf{Q})}^{q}
 +\left\Vert \int_{0}^{T}\bigg(\sum_{n=1}^{N}Y_{t}^{(n)}\bigg)dt
 \right\Vert _{L^{q}(\mathbf{Q})}^{\frac{q}{2}}\right)\nonumber \\
 & \overset{(\ref{eq:Hoelder-ungl.})}{\leq} & 
 \const\left(\mathbf{E_{Q}}\left[\int_{0}^{T}\bigg|\sum_{m=1}^{M}X_{t}^{(m)}\bigg|^{q}dt\right]
 +\sqrt{\mathbf{E_{Q}}\left[\int_{0}^{T}
 \bigg|\sum_{n=1}^{N}Y_{t}^{(n)}\bigg|^{q}dt\right]}\right)\nonumber \\
 & \overset{(\ref{eq:convex-q})}{\leq} & 
 \const\left(\sum_{m=1}^{M}\mathbf{E_{Q}}\left[\int_{0}^{T}|X_{t}^{(m)}|^{q}dt\right]
 +\sqrt{\sum_{n=1}^{N}\mathbf{E_{Q}}\left[\int_{0}^{T}|Y_{t}^{(n)}|^{q}dt\right]}\right)
 \label{eq:Summen-Norm}
\end{eqnarray}
for any $M,N\in\mathbb{N}$ and processes
$X^{(m)}$, $m=1,\dots,M$ and
$Y^{(n)}$, $n=1,\dots,N$.
\end{enumerate}
\end{remark}

\begin{lemma}
Assume Condition (\ref{eq:Bed.Primal}).
\begin{enumerate}
 \item For any stopping time $\tau$ we have
\begin{equation}
\mathbf{E_{Q}}\left[X_\tau^{\varphi^{\varepsilon},\varepsilon}-(x+\varphi\mal S_\tau)\right]
=p\mathbf{E_{Q}}\left[\big(2\Delta\varphi^{2}-(\Delta\varphi^{+})^{2}\big)\mal[S,S]_\tau\right]
+O(\varepsilon),\label{eq:Delta-X_1}
\end{equation}
\begin{equation}
\mathbf{E_{Q}}\left[\big(X_\tau^{\varphi^{\varepsilon},\varepsilon}-(x+\varphi\mal S_\tau)\big)^{2}\right]
=\mathbf{E_{Q}}\left[\Delta\varphi^{2}\mal[S,S]_\tau\right]+O(\varepsilon),\label{eq:Delta-X_2}
\end{equation}
\begin{equation}
\left\Vert X^{\varphi^{\varepsilon},\varepsilon}-(x+\varphi\mal S)\right\Vert _{S^{3}(\mathbf{Q})}^{3}
=O(\varepsilon).\label{eq:Delta-X_3}
\end{equation}
\item Define stopping times
\begin{equation}
\tau^{\varepsilon,1}:=\inf\left\{ t\in[0,T]:
\big|X_{t}^{\varphi^{\varepsilon},\varepsilon}-(x+\varphi\mal S_{t})\big|>1\right\} ,\label{eq:stop_tao_1}
\end{equation}
\begin{equation}
\tau^{\varepsilon,2}:=\inf\left\{ t\in[0,T]:
\big|X_{t}^{\varphi^{\varepsilon},\varepsilon}\big|>\varepsilon^{-4/3}\right\} ,\label{eq:stop_tao_2}
\end{equation}
\begin{equation}
\tau^{\varepsilon}:=\tau^{\varepsilon,1}\wedge\tau^{\varepsilon,2}\wedge T.\label{eq:stop_tao}
\end{equation}
Then
\begin{equation}
\mathbf{Q}(\tau^{\varepsilon}<T)=O\big(\varepsilon^{4/3}\big)\label{eq:Ws(Tao<=00003DT)}
\end{equation}
and 
$\lim_{\varepsilon\downarrow0}\mathbf{P}(\tau^{\varepsilon}<T)=0$.
\end{enumerate}
\end{lemma}

\begin{prf}
\emph{1.} Note that
\begin{eqnarray}
X^{\varphi^{\varepsilon},\varepsilon}-(x+\varphi\mal S) 
& = & (\varphi^{\varepsilon}\mal S^{\varepsilon}-\varphi\mal S)
+X^{\varphi^{\varepsilon},\varepsilon}-(x+\varphi^{\varepsilon}\mal S^{\varepsilon})\nonumber\\
 & = & \Delta\varphi\mal S+\varphi\mal\Delta S+\Delta\varphi\mal\Delta S
 +\big(X^{\varphi^{\varepsilon},\varepsilon}-(x+\varphi^{\varepsilon}\mal S^{\varepsilon})\big).
 \label{e:notethat}
\end{eqnarray}
From (\ref{eq:S-Norm<=00003DH-Norm}, \ref{eq:Summen-Norm})
and Condition (\ref{eq:Bed.Primal}), we get
\begin{equation}
\Vert \Delta\varphi\mal S\Vert _{S^{q}(\mathbf{Q})}^q
\leq \const\Vert \Delta\varphi\mal S\Vert _{H^{q}(\mathbf{Q})}^{q}
\leq \const\underbrace{\sqrt{\mathbf{E_{Q}}\left[\int_{0}^{T}\big|\beta_{t}^{2}
c_{t}^{S,S}\big|^{q}dt\right]}}_{<\infty}\varepsilon^{\frac{q}{3}}\label{eq:X-1_3-6}
\end{equation}
for $q\leq4$.
By letting $q=2$ we deduce that $\Delta\varphi\mal S$
is a square-integrable $\mathbf{Q}$-martingale. Thus
\begin{equation}
\mathbf{E_{Q}}[\Delta\varphi\mal S_\tau]=0,\label{eq:X-1_1}
\end{equation}
\begin{equation}
\mathbf{E_{Q}}\left[(\Delta\varphi\mal S_\tau)^{2}\right]
=\mathbf{E_{Q}}\left[\Delta\varphi^{2}\mal[S,S]_\tau\right]\label{eq:X-1_2}
\end{equation}
for any
stopping time $\tau$.
Integration by parts yields
\[
\varphi\mal\Delta S=\varphi\Delta S-\Delta S\mal\varphi-[\Delta S,\varphi].
\]
By $|\Delta S_t|\leq \varepsilon|S|$ and Condition (\ref{eq:Bed.Primal}), we have
\begin{equation}
\Vert \varphi\Delta S\Vert _{S^{q}(\mathbf{Q})}^{q}\leq
\underbrace{\mathbf{E_{Q}}\left[\sup_{t\in[0,T]}|\varphi_{t}S_{t}|^{q}\right]}_{<\infty}
\varepsilon^{q}\label{eq:X-2a_1-6}
\end{equation}
for $q\leq 4$.
From $|\Delta S_t|\leq \varepsilon|S|$, (\ref{eq:Summen-Norm}), and Condition (\ref{eq:Bed.Primal}), we obtain
\begin{equation}
\Vert \Delta S\mal\varphi\Vert _{H^{q}(\mathbf{Q})}^{q}
\leq \const\underbrace{\left(\mathbf{E_{Q}}\left[\int_{0}^{T}|S_{t}b_{t}^{\varphi}|^{q}dt\right]
+\sqrt{\mathbf{E_{Q}}\left[\int_{0}^{T}|S_{t}^{2}c_{t}^{\varphi,\varphi}|^qdt\right]}\right)}_{<\infty}
\varepsilon^{q}\label{eq:X-2b_1-6}
\end{equation}
for $q\leq 4$.

In view of (\ref{eq:c_delta-S,phi}, \ref{eq:c_delta-S,phi<=00003D}, \ref{eq:Summen-Norm}),
and Condition (\ref{eq:Bed.Primal}), 
\begin{equation}
\mathbf{E_{Q}}[[\Delta S,\varphi]_\tau]=-p\mathbf{E_{Q}}\left[\big(\Delta\varphi^{2
}-(\Delta\varphi^{+})^{2}\big)\mal[S,S]_\tau\right]+O(\varepsilon),\label{eq:X-2c_1}
\end{equation}
\begin{equation}
\left\Vert [\Delta S,\varphi]\right\Vert _{H^{q}(\mathbf{Q})}^{q}\leq 
\const\underbrace{\sum_{c^{\Delta,\varphi}\in\mathcal{H}^{c^{\Delta S,\varphi}}}
\mathbf{E_{Q}}\left[\int_{0}^{T}|c_{t}^{\Delta,\varphi}|^{q}dt\right]}_{<\infty}
\varepsilon^{\frac{2q}{3}}
\label{eq:X-2c_2-6}
\end{equation}
holds for any stopping time $\tau$ and $q\leq4$.
For any stopping time $\tau$ and  $q\leq4$, (\ref{eq:X-2a_1-6}--\ref{eq:X-2c_2-6}) then yield
\begin{equation}
\mathbf{E_{Q}}[\varphi\mal\Delta S_\tau]=
p\mathbf{E_{Q}}\left[\big(\Delta\varphi^{2}-(\Delta\varphi^{+})^{2}\big)\mal[S,S]_\tau\right]
+O(\varepsilon),\label{eq:X-2_1}
\end{equation}
\begin{equation}
\Vert \varphi\mal\Delta S\Vert _{S^{q}(\mathbf{Q})}^{q}=O\big(\varepsilon^{{2q}/{3}}\big).
\label{eq:X-2_2-6}
\end{equation}
From (\ref{eq:b_delta-S}, \ref{eq:b_delta-S <=00003D}, \ref{eq:c_delta-S}, \ref{eq:Summen-Norm}), 
and Condition (\ref{eq:Bed.Primal}),
we obtain 
\begin{equation}
\mathbf{E_{Q}}[\Delta\varphi\mal\Delta S_\tau]
=p\mathbf{E_{Q}}\left[\Delta\varphi^{2}\mal[S,S]_\tau\right]+O(\varepsilon),\label{eq:X-3_1}
\end{equation}
\begin{eqnarray}
\Vert \Delta\varphi\mal\Delta S\Vert _{H^{q}(\mathbf{Q})}^{q} & \leq & 
\const\underbrace{\sum_{b^{\Delta}\in\mathcal{H}^{b^{\Delta S}}}
\mathbf{E_{Q}}\left[\int_{0}^{T}|\beta_{t}b_{t}^{\Delta}|^{q}dt\right]}_{<\infty}
\varepsilon^{\frac{2q}{3}}\nonumber \\
 &  & {}+\const\underbrace{\sqrt{\sum_{c^{\Delta,\Delta}\in\mathcal{H}^{c^{\Delta S,\Delta S}}}
 \mathbf{E_{Q}}\left[\int_{0}^{T}|\beta_{t}^{2}c_{t}^{\Delta,\Delta}|^{q}dt\right]}}_{<\infty}
 \varepsilon^{\frac{2q}{3}}\label{eq:X-3_2-6}
\end{eqnarray}
for any stopping time $\tau$ and $q\leq4$.
Due to Remark~\ref{rem:self-fin} and  (\ref{eq:Rand_delta-S_+}, \ref{eq:Rand_delta-S_-}),
\begin{eqnarray}
\left|X_{t}^{\varphi^{\varepsilon},\varepsilon}-(x+\varphi^{\varepsilon}\mal S_{t}^{\varepsilon})\right| 
& \leq & 2|\varphi_{t}^{\varepsilon}S_{t}|\varepsilon+x^{S}\varepsilon\nonumber \\
 & \leq & \const\left(|\varphi_{t}S_{t}|+|\beta_{t}S_{t}|+x^{S}\right)\varepsilon\label{eq:liq<=00003D}
\end{eqnarray}
holds for any $t\in[0,T]$.
So by (\ref{eq:3-Eks-Ungl.}) and Condition (\ref{eq:Bed.Primal}), we obtain
\begin{equation}
\left\Vert X^{\varphi^{\varepsilon},\varepsilon}
-(x+\varphi^{\varepsilon}\mal S^{\varepsilon})\right\Vert _{S^{q}(\mathbf{Q})}^{q}
\leq \const\underbrace{\mathbf{E_{Q}}\left[\sup_{t}|\varphi_{t}S_{t}|^{q}
+\sup_{t}|\beta_{t}S_{t}|^{q}+x^{S}\right]}_{<\infty}\varepsilon^{q}\label{eq:X-liq_1-6}
\end{equation}
for $q\leq4$.
In view of (\ref{eq:3-Eks-Ungl.}), relations 
(\ref{e:notethat}) and
(\ref{eq:X-1_1}, \ref{eq:X-2_1}, \ref{eq:X-3_1}, \ref{eq:X-liq_1-6}) for $q=1$ imply (\ref{eq:Delta-X_1}).
(\ref{e:notethat}) and (\ref{eq:X-1_2}, \ref{eq:X-2_2-6}, \ref{eq:X-3_2-6}, \ref{eq:X-liq_1-6})
for $q=2$ yield (\ref{eq:Delta-X_2}).
(\ref{e:notethat}) and (\ref{eq:X-1_3-6}, \ref{eq:X-2_2-6}, \ref{eq:X-3_2-6}, \ref{eq:X-liq_1-6})
for $q=3$ imply (\ref{eq:Delta-X_3}).

\emph{2.}
(\ref{eq:stop_tao_1}), Markov's inequality, 
and (\ref{eq:3-Eks-Ungl.}, \ref{e:notethat}, \ref{eq:X-1_3-6}, \ref{eq:X-2_2-6}, \ref{eq:X-3_2-6}, \ref{eq:X-liq_1-6})  yield
\begin{eqnarray*}
 \mathbf{Q}(\tau^{\varepsilon,1}< T)
&\leq&\mathbf{Q}\left(\sup_{t\in[0,T]}\big|X_t^{\varphi^{\varepsilon},\varepsilon}-(x+\varphi\mal S_t)\big|>1\right)\\
&\leq&\left\Vert X^{\varphi^{\varepsilon},\varepsilon}
-(x+\varphi\mal S)\right\Vert _{S^{4}(\mathbf{Q})}^{4}\\
&=&O\big(\varepsilon^{4/3}\big).
\end{eqnarray*}
Condition (\ref{eq:Bed.U_2}) implies that the $\mathbf{Q}$-martingale $\varphi\mal S$ is in fact 
square-integrable and hence
$\mathbf{E_{Q}}[\varphi^2\mal[S,S]_T]<\infty$.
Moreover, from (\ref{eq:stop_tao_2}), Markov's inequality, (\ref{eq:S-Norm<=00003DH-Norm}, \ref{e:notethat}),
and (\ref{eq:Summen-Norm}, \ref{eq:X-1_3-6}, \ref{eq:X-2_2-6}, \ref{eq:X-3_2-6}, \ref{eq:X-liq_1-6}) for
$q=1$, we get
\begin{eqnarray*}
\mathbf{Q}(\tau^{\varepsilon,2}< T) 
&\leq&\mathbf{Q}\left(\sup_{t\in[0,T]}\big|X_t^{\varphi^{\varepsilon},\varepsilon}\big|
>\varepsilon^{-4/3}\right)\\
& \leq & \big\Vert X^{\varphi^{\varepsilon}}\big\Vert _{S^{1}(\mathbf{Q})}\varepsilon^{4/3}\\
& \leq & \left(\Vert x+\varphi\mal S\Vert _{S^{1}(\mathbf{Q})}
+\big\Vert X^{\varphi^{\varepsilon},\varepsilon}-(x+\varphi\mal S)\big\Vert _{S^{1}
(\mathbf{Q})}\right)\varepsilon^{4/3}\\
& \leq & \const\underbrace{\left(1+\sqrt{\mathbf{E_{Q}}\left[\varphi^{2}\mal[S,S]_{T}\right]}
+\big\Vert X^{\varphi^{\varepsilon},\varepsilon}-(x+\varphi\mal S)\big\Vert _{S^{1}
(\mathbf{Q})}\right)}_{<\infty}\varepsilon^{4/3}.
\end{eqnarray*}
Therefore, 
$\mathbf{Q}(\tau^{\varepsilon}< T)\leq\mathbf{Q}(\tau^{\varepsilon,1}< T)
+\mathbf{Q}(\tau^{\varepsilon,2}< T)=O(\varepsilon^{4/3})$.
Since $\mathbf{P}$ and $\mathbf{Q}$ are equivalent,
$\mathbf{Q}(\tau^{\varepsilon}< T)\to1$ implies
$\mathbf{P}(\tau^{\varepsilon}< T)\to1$.
\end{prf}

\begin{lemma}\label{lem:Primal}
Suppose that Conditions~(\ref{eq:Bed.Primal}, \ref{eq:Bed.U_2}) hold.
For $\tau^{\varepsilon}$ as in (\ref{e:tauepsilon}) resp.\  (\ref{eq:stop_tao}), we have
\begin{eqnarray}
\lefteqn{\mathbf{E}\left[\uti\big(X_{\tau^{\varepsilon}}^{\varphi^{\varepsilon},\varepsilon}\big)\right]}\nonumber\\
 & = & \mathbf{E}[\uti(x+\varphi\mal S_{T})]
 -yp\mathbf{E_{Q}}\left[\Big((\Delta\varphi^{+})^{2}-\frac{3}{2}\Delta\varphi^{2}\Big)\mal[S,S]_{T}\right]
 +O(\varepsilon).\label{e:imlemma}
\end{eqnarray}
\end{lemma}

\begin{prf}
Let $Z$ denote the density process of $\mathbf{Q}$.
Taylor expansion of $\uti(x)=-e^{-px}$ yields
\begin{eqnarray*}
 \lefteqn{ \uti\big(X_{\tau^{\varepsilon}}^{\varphi^{\varepsilon},\varepsilon}\big)}\\
 & = & \uti(x+\varphi\mal S_{T})\\
 &  & {}+\underbrace{yZ_{T}\big(X_{\tau^{\varepsilon}}^{\varphi^{\varepsilon},\varepsilon}
 -(x+\varphi\mal S_{T})\big)}_{=:G^{(5)}}-\underbrace{\frac{p}{2}y
 Z_{T}\big(X_{\tau^{\varepsilon}}^{\varphi^{\varepsilon},\varepsilon}-
 (x+\varphi\mal S_{T})\big)^{2}}_{=:G^{(6)}}\\
 &  & {} +\underbrace{\frac{p^{2}}{6}yZ_{T}
 \exp\Big(-p\theta\big(X_{\tau^{\varepsilon}}^{\varphi^{\varepsilon},\varepsilon}
 -(x+\varphi\mal S_{T})\big)\Big)\big(X_{\tau^{\varepsilon}}^{\varphi^{\varepsilon},\varepsilon}
 -(x+\varphi\mal S_{T})\big)^{3}}_{=:G^{(7)}}
\end{eqnarray*}
for some random $\theta\in(0,1)$.
Notice that
\begin{equation}
X_{\tau^{\varepsilon}}^{\varphi^{\varepsilon},\varepsilon}
-(x+\varphi\mal S_{T})=\big(X_{\tau^{\varepsilon}}^{\varphi^{\varepsilon},\varepsilon}
-(x+\varphi\mal S_{\tau^{\varepsilon}})\big)+(\varphi\mal S_{\tau^{\varepsilon}}
-\varphi\mal S_{T})\label{eq:X=00003DTao+(T-Tao)}
\end{equation}
with
\begin{equation}
\left|X_{\tau^{\varepsilon}}^{\varphi^{\varepsilon},\varepsilon}
-(x+\varphi\mal S_{\tau^{\varepsilon}})\right|\overset{(\ref{eq:stop_tao_1})}{\leq}1,\label{eq:X-Tao besch.}
\end{equation}
and
\begin{eqnarray}
\mathbf{E_{Q}}\left[|\varphi\mal S_{\tau^{\varepsilon}}-\varphi\mal S_{T}|^{2n}\right]
&\leq&\const\left(\mathbf{E_{Q}}\left[|\varphi\mal S_{\tau^{\varepsilon}}|^{2n}\right]
+\mathbf{E_{Q}}\left[|\varphi\mal S_{T}|^{2n}\right]\right)\nonumber\\
&\leq&\const\mathbf{E_{Q}}\left[|\varphi\mal S_{T}|^{2n}\right]
\overset{(\ref{eq:Bed.U_2})}{<}\infty\label{eq:X-(Tao-T) sub-}
\end{eqnarray}
for $n\in\mathbb{N}$, where the first inequality is due to the fact
that $|\varphi\mal S|^{2n}$ is a $\mathbf{Q}$-submartingale.
In view of (\ref{eq:X=00003DTao+(T-Tao)}), Hölder's inequality, and
(\ref{eq:Ws(Tao<=00003DT)}, \ref{eq:X-(Tao-T) sub-}, \ref{eq:X-Tao besch.}), we have
\begin{eqnarray}
\lefteqn{\left|\mathbf{E}[G^{(5)}]-y
\mathbf{E_{Q}}\left[X_{\tau^{\varepsilon}}^{\varphi^{\varepsilon},\varepsilon}
-(x+\varphi\mal S_{\tau^{\varepsilon}})\right]\right|}\nonumber \\
 & \leq & y\mathbf{E_{Q}}\left[\mathbf{1}_{\{ \tau^{\varepsilon}< T\} }
 |\varphi\mal S_{\tau^{\varepsilon}}-\varphi\mal S_{T}|\right]\nonumber \\
 & \leq & y\mathbf{Q}(\tau^{\varepsilon}< T)^{\frac{3}{4}}
 \sqrt[4]{\mathbf{E_{Q}}\left[|\varphi\mal S_{\tau^{\varepsilon}}
 -\varphi\mal S_{T}|^{4}\right]}\nonumber \\
 & = & O(\varepsilon),\label{eq:X-(Tao-T)_1 klein}
\end{eqnarray}
and
\begin{eqnarray}
\lefteqn{ \left|\mathbf{E}[G^{(6)}]-\frac{p}{2}
y\mathbf{E_{Q}}\left[\big(X_{\tau^{\varepsilon}}^{\varphi^{\varepsilon},\varepsilon}
-(x+\varphi\mal S_{\tau^{\varepsilon}})\big)^{2}\right]\right|}\nonumber \\ 
& \leq & \const
\left(\mathbf{E_{Q}}\left[\mathbf{1}_{\{ \tau^{\varepsilon}< T\} }
|\varphi\mal S_{\tau^{\varepsilon}}-\varphi\mal S_{T}|^{2}\right]
+\mathbf{E_{Q}}\left[\mathbf{1}_{\{ \tau^{\varepsilon}< T\} }
 |\varphi\mal S_{\tau^{\varepsilon}}-\varphi\mal S_{T}|\right]\right)
\nonumber \\
 & \leq & \const\mathbf{Q}(\tau^{\varepsilon}< T)^{\frac{3}{4}}
 \sqrt[4]{\mathbf{E_{Q}}\left[|\varphi\mal S_{\tau^{\varepsilon}}
 -\varphi\mal S_{T}|^{8}\right]}+O(\varepsilon)\nonumber \\
 & = & O(\varepsilon).\label{eq:X-(Tao-T)_2 klein}
\end{eqnarray}
From (\ref{eq:X-Tao besch.}) and (\ref{eq:Delta-X_3}), we obtain
\begin{eqnarray*}
 \lefteqn{ |\mathbf{E}[G^{(7)}]|}\\
 & \leq & \const\mathbf{E_{Q}}\left[\exp\big(-p\theta(\varphi\mal S_{\tau^{\varepsilon}}
 -\varphi\mal S_{T})\big)\big|X_{\tau^{\varepsilon}}^{\varphi^{\varepsilon},\varepsilon}
 -(x+\varphi\mal S_{T})\big|^{3}\right]\\
 & \leq & \const\underbrace{\mathbf{E_{Q}}\left[\mathbf{1}_{\{ \tau^{\varepsilon}=T\} }
 \big|X_{T}^{\varphi^{\varepsilon},\varepsilon}-(x+\varphi\mal S_{T})\big|^{3}\right]}_{=O(\varepsilon)}\\
 &  & {}+\const\underbrace{\mathbf{E_{Q}}\left[\mathbf{1}_{\{ \tau^{\varepsilon}< T\} }
 \exp\big(-p\theta(\varphi\mal S_{\tau^{\varepsilon}}-\varphi\mal S_{T})\big)
 \big|X_{\tau^{\varepsilon}}^{\varphi^{\varepsilon},\varepsilon}-(x+\varphi\mal S_{T})\big|^{3}\right]}_{=:G^{(8)}}.
\end{eqnarray*}
By Hölder's inequality and (\ref{eq:Ws(Tao<=00003DT)}) we have
\[
G^{(8)}\leq\underbrace{\mathbf{Q}(\tau^{\varepsilon}< T)^{\frac{3}{4}}}_{=O(\varepsilon)}G^{(9)}
\]
with
\[
G^{(9)}=\sqrt[4]{\mathbf{E_{Q}}\left[\exp\big(-4p\theta(\varphi\mal S_{\tau^{\varepsilon}}
-\varphi\mal S_{T})\big)\big|X_{\tau^{\varepsilon}}^{\varphi^{\varepsilon},\varepsilon}
-(x+\varphi\mal S_{T})\big|^{12}\right]}.
\]
Again by Hölder's inequality,
\[
G^{(9)}\leq G^{(10)}\underbrace{\sqrt[36]{\mathbf{E_{Q}}
\left[\big|X_{\tau^{\varepsilon}}^{\varphi^{\varepsilon},
\varepsilon}-(x+\varphi\mal S_{T})\big|^{108}\right]}}_{\overset{(\ref{eq:X-(Tao-T) sub-})}{<}\infty}
\]
with
\begin{eqnarray*}
G^{(10)} & = & \sqrt[9/2]{\mathbf{E_{Q}}\left[\exp\Big(-\frac{9}{2}
p\theta(\varphi\mal S_{\tau^{\varepsilon}}-\varphi\mal S_{T})\Big)\right]}\\
 & \leq & \sqrt[9]{\mathbf{E_{Q}}\left[\exp(-9p\theta\varphi\mal S_{\tau^{\varepsilon}})\right]}
 \underbrace{\sqrt[9]{\mathbf{E_{Q}}\left[
 \exp(9p\theta\varphi\mal S_{T})\right]}}_{\overset{(\ref{eq:Bed.U_2})}{<}\infty}.
\end{eqnarray*}
Since $\varphi\mal S$ is a $\mathbf{Q}$-martingale,
$\exp(-9p\theta\varphi\mal S)$
is a $\mathbf{Q}$-submartingale by Jensen's inequality. Hence
\[
\mathbf{E_{Q}}[\exp(-9p\theta\varphi\mal S_{\tau^{\varepsilon}})]
\leq\mathbf{E_{Q}}[\exp(-9p\theta\varphi\mal S_{T})]\overset{(\ref{eq:Bed.U_2})}{<}\infty.
\]
Therefore
\begin{equation}
\mathbf{E}[G^{(7)}]=O(\varepsilon).\label{eq:Rest<Delta-X_3 klein}
\end{equation}

Combining (\ref{eq:X-(Tao-T)_1 klein}, \ref{eq:X-(Tao-T)_2 klein}, \ref{eq:Rest<Delta-X_3 klein}) 
with (\ref{eq:Delta-X_1}, \ref{eq:Delta-X_2}), we obtain
\begin{eqnarray*}
\lefteqn{\mathbf{E}\left[\uti\big(X_{\tau^{\varepsilon}}^{\varphi^{\varepsilon},\varepsilon}\big)\right]}\\
 & = & \mathbf{E}\left[\uti(x+\varphi\mal S_{T})\right]
 -yp\mathbf{E_{Q}}\left[\Big((\Delta\varphi^{+})^{2}
 -\frac{3}{2}\Delta\varphi^{2}\Big)\mal[S,S]_{\tau^{\varepsilon}}\right]+O(\varepsilon).
\end{eqnarray*}
Moreover, Hölder's inequality and (\ref{eq:Ws(Tao<=00003DT)}) yield
\begin{eqnarray}
 \lefteqn{ \left|\mathbf{E}_\mathbf{Q}\left[\int_{\tau^{\varepsilon}}^{T}
 \Big(	(\Delta\varphi_{t}^{+})^{2}
 -\frac{3}{2}\Delta\varphi_{t}^{2}\Big)d[S,S]_{t}\right]\right|}\nonumber \\
 & \leq & \const\underbrace{\mathbf{Q}(\tau^{\varepsilon}< T)^{\frac{3}{4}}}_{=O(\varepsilon)}
 \underbrace{\sqrt[4]{\mathbf{E_{Q}}\left[((\Delta\varphi^{+})^{2}\mal[S,S]_{T})^{4}\right]}}_{=:G^{(11)}},
 \label{eq:Verlust(Tao-T) klein}
\end{eqnarray}
where
\[
G^{(11)}\overset{(\ref{eq:Hoelder-ungl.})}
{\leq}\const\underbrace{\sqrt[4]{\mathbf{E_{Q}}
\left[\int_{0}^{T}|\beta_{t}^{2}c_{t}^{S,S}|^{4}dt\right]}}_{\overset{(\ref{eq:Bed.Primal})}{<}\infty}
\varepsilon^{2/3}.
\]
This completes the proof.
\end{prf}

\subsection{Dual considerations}
In order to obtain an approximate upper bound to the maximal expected utility,
we construct a dual variable based
on Girsanov's theorem. More specifically, we consider the minimal martingale measure
for the appropriately stopped process $S^\varepsilon$ relative to $\mathbf{Q}$.
This martingale measure turns out to be optimal to the leading order.

Let
\begin{equation}
 \rho^{\varepsilon,1}
 :=\inf\left\{ t\in[0,T]:\left|\frac{c_{t}^{\Delta S,S}}{c_{t}^{S,S}}\right|>\frac{1}{2}\right\} 
\label{eq:stop_sigma}
\end{equation}
and define
$Z^{\varepsilon,\mathbf{Q}}:=\exp(N^{\varepsilon})$
with
\begin{equation}
N^{\varepsilon}:=-\int_{0}^{\cdot}\theta_{t}^{\varepsilon}dS_{t}-
\frac{1}{2}\int_{0}^{\cdot}(\theta_{t}^{\varepsilon})^{2}d[S,S]_{t},\label{eq:Def.Z}
\end{equation}
where
\begin{equation}
\theta^{\varepsilon}:=\frac{b^{S^{\varepsilon}}}{c^{S^{\varepsilon},S}}
\mathbf{1}_{\auf0,\rho^{\varepsilon,1}\zu}
=\frac{b^{\Delta S}}{c^{S,S}+c^{\Delta S,S}}
\mathbf{1}_{\auf0,\rho^{\varepsilon,1}\zu}.\label{eq:Def.Theta}
\end{equation}
Furthermore, let
\begin{eqnarray}
\rho^{\varepsilon,2}&:=&\inf\left\{ t\in[0,T]:|Z_{t}^{\varepsilon,\mathbf{Q}}-1|>\frac{1}{2}\right\},
\label{eq:stop_Z}\\
\rho^{\varepsilon}&:=&\rho^{\varepsilon,1}\wedge\rho^{\varepsilon,2}\wedge T\label{eq:stop_rho}
\end{eqnarray}
and define the ``stopped'' processes
\[
\overline{\varphi}^{\varepsilon}:=\varphi^{\varepsilon}
\mathbf{1}_{\auf0,\tau^{\varepsilon}\wedge\rho^{\varepsilon}\zu},
\quad\Delta\overline{\varphi}^{\varepsilon}:=\Delta\varphi^{\varepsilon}
\mathbf{1}_{\auf0,\tau^{\varepsilon}\wedge\rho^{\varepsilon}\zu},
\]
\begin{equation}
\quad\overline{S}^{\varepsilon}
:=S\left(1+\frac{\Delta S^{\tau^{\varepsilon}\wedge\rho^{\varepsilon}}}
{S^{\tau^{\varepsilon}\wedge\rho^{\varepsilon}}}\right),
\quad\overline{N}^{\varepsilon}:=
(N^{\varepsilon})^{\tau^{\varepsilon}\wedge\rho^{\varepsilon}},
\quad\overline{Z}^{\varepsilon}:=
(Z^{\varepsilon,\mathbf{Q}})^{\tau^{\varepsilon}\wedge\rho^{\varepsilon}}.\label{eq:Def.stop_phi, S, Z}
\end{equation}

\begin{remark}\label{rem:Dual-EMM}
By construction, $\overline{Z}^{\varepsilon}$ is a bounded $\mathbf{Q}$-local
martingale and hence a $\mathbf{Q}$-martingale.
If $Z$ denotes the density process of $\mathbf{Q}$,
the process $Z^{\varepsilon}:=Z\overline{Z}^{\varepsilon}$ is a $\mathbf{P}$-martingale.
Integration by parts yields that $\overline{Z}^{\varepsilon}\overline{S}^{\varepsilon}$
is a $\mathbf{Q}$-local martingale and hence 
$Z^{\varepsilon}\overline{S}^{\varepsilon}=Z\overline{Z}^{\varepsilon}\overline{S}^{\varepsilon}$
is a $\mathbf{P}$-local martingale. Consequently, $\overline{S}^{\varepsilon}$
a local martingale under the probability measure with density process
$Z^{\varepsilon}$. Therefore, $Z^{\varepsilon}$ corresponds
to an equivalent (local) martingale measure for $\overline{S}^{\varepsilon}$.
It serves as a dual variable in the frictionless market with shadow
price $\overline{S}^{\varepsilon}$.
\end{remark}

\begin{lemma}
\begin{enumerate}
 \item  Conditions (\ref{eq:Bed.Dual_sigma}, \ref{eq:Bed.Dual}) imply
\begin{equation}
\mathbf{E_{Q}}\left[(\overline{Z}_{T}^{\varepsilon}-1)^{2}\right]
=p^{2}\mathbf{E_{Q}}\left[\Delta\overline{\varphi}^{2}\mal[S,S]_{T}\right]
+O(\varepsilon)
\label{eq:Delta-Z_2}
\end{equation}
and
\begin{equation}
\mathbf{E_{Q}}\left[|\overline{Z}_{T}^{\varepsilon}-1|^{3}\right]
=O(\varepsilon).\label{eq:Delta-Z_3}
\end{equation}
\item Condition (\ref{eq:Bed.Dual_sigma}) implies
\begin{equation}
\mathbf{Q}(\rho^{\varepsilon,1}< T)=O\big(\varepsilon^{4/3}\big).\label{eq:Ws(rho_sigma<T)}
\end{equation}
\item Conditions (\ref{eq:Bed.Dual_sigma}, \ref{eq:Bed.Dual}) imply
\begin{equation}
\mathbf{Q}(\rho^{\varepsilon,2}< T)=O\big(\varepsilon^{4/3}\big).\label{eq:Ws(rho_Z<T)}
\end{equation}
\end{enumerate}
\end{lemma}

\begin{prf}
Note that on $\{\rho^{\varepsilon}\geq t\} $, 
(\ref{eq:b_delta-S}, \ref{eq:b_delta-S <=00003D}, \ref{eq:sigma_delta-S<=00003D}, 
 \ref{eq:stop_sigma}, \ref{eq:Def.Theta})
yield
\begin{equation}
\theta_{t}^{\varepsilon}\sigma_{t}^{S}=p\Delta\varphi_{t}\sigma_{t}^{S}
+\underbrace{\frac{b_{t}^{\Delta S}\sigma_{t}^{S}}{c_{t}^{S,S}
+c_{t}^{\Delta S,S}}-p\Delta\varphi_{t}\sigma_{t}^{S}}_{=:G^{(12)}_{t}},\label{eq:Theta}
\end{equation}
\begin{equation}
|\theta_{t}^{\varepsilon}\sigma_{t}^{S}|
\leq  \const\sum_{b^{\Delta}\in\mathcal{H}^{b^{\Delta S}}}
\left|\frac{b_{t}^{\Delta}}{\sigma_{t}^{S}}\right|\varepsilon^{1/3},
\label{e:neudazu}
\quad
|p\Delta\varphi\sigma^{S}|\leq \const\beta\sigma^{S}
\varepsilon^{1/3},
\end{equation}
\begin{equation}
\big|G^{(12)}_{t}\big|
\leq \const\sum_{b^{\Delta}\in\mathcal{H}^{b^{\Delta S}}}
\left|\frac{b_{t}^{\Delta}}{\sigma_{t}^{S}}\right|
\left(1+\sum_{c^{\Delta,S}\in\mathcal{H}^{c^{\Delta S,S}}}
\left|\frac{c_{t}^{\Delta,S}}{c^{S,S}}\right|\right)
\varepsilon^{2/3}.\label{eq:Theta<=00003D}
\end{equation}
From (\ref{eq:stop_Z}) and Taylor expansion of $x\mapsto e^{x}$, we obtain
\begin{equation}
\overline{Z}^{\varepsilon}-1=-p\Delta\overline{\varphi}\mal S+(\overline{N}^{\varepsilon}
+p\Delta\overline{\varphi}\mal S)+G^{(13)}\quad\textrm{with}\quad\big|G^{(13)}\big|
\leq\frac{3}{4}|\overline{N}^{\varepsilon}|^{2}.\label{eq:delta-Z<=00003D}
\end{equation}

\emph{1.} From (\ref{eq:delta-Z<=00003D}), (\ref{eq:3-Eks-Ungl.}) for
$q=2$, (\ref{eq:S-Norm<=00003DH-Norm}) for $q=2,4$, 
(\ref{eq:Theta}, \ref{e:neudazu}, \ref{eq:Theta<=00003D}, \ref{eq:Summen-Norm}),
Cauchy-Schwarz, and Conditions (\ref{eq:Bed.Dual_sigma}, \ref{eq:Bed.Dual}), we deduce that
\begin{eqnarray}
\lefteqn{\mathbf{E_{Q}}\left[(\overline{Z}_{T}^{\varepsilon}-1
+p\Delta\overline{\varphi}\mal S_{T})^{2}\right]}\nonumber \\
 & \leq & \const\Vert \overline{N}^{\varepsilon}+p\Delta\overline{\varphi}\mal S
 \Vert _{H^{2}(\mathbf{Q})}^{2}+\const\Vert \overline{N}^{\varepsilon}
 \Vert _{H^{4}(\mathbf{Q})}^{4}\nonumber \\
 & \leq & \const\sum_{m=1,2}\sum_{n=1,2}\underbrace{
 \left(\sum_{b^{\Delta}\in\mathcal{H}^{b^{\Delta S}}}
 \mathbf{E_{Q}}\left[\int_{0}^{T}\bigg|\frac{b_{t}^{\Delta}}
 {\sigma_{t}^{S}}\bigg|^{8m}dt\right]\right)^{\frac{1}{2n}}}_{<\infty}\nonumber\\ 
&&{}\times\underbrace{ \left(1+\sum_{c^{\Delta,S}\in\mathcal{H}^{c^{\Delta S,S}}}
 \mathbf{E_{Q}}\left[\int_{0}^{T}\bigg|\frac{c_{t}^{\Delta,S}}
 {c_{t}^{S,S}}\bigg|^{8m}dt\right]\right)^{\frac{1}{2n}}}_{<\infty}
 \varepsilon^{4/3}.\label{eq:Delta-Z_2'}
\end{eqnarray}
Combined with (\ref{eq:X-1_2}) and Cauchy-Schwarz' inequality, this
yields (\ref{eq:Delta-Z_2}).
Analogously, from (\ref{eq:delta-Z<=00003D}), (\ref{eq:3-Eks-Ungl.})
for $q=3$, (\ref{eq:S-Norm<=00003DH-Norm}) for $q=3,6$, 
(\ref{eq:Theta}, \ref{e:neudazu}, \ref{eq:Summen-Norm}),
and Condition (\ref{eq:Bed.Dual}), we get
\begin{eqnarray*}
\mathbf{E_{Q}}\left[|\overline{Z}_{T}^{\varepsilon}-1|^{3}\right] 
& \leq & \const\Vert \overline{N}^{\varepsilon}\Vert _{H^{3}(\mathbf{Q})}^{3}
+\const\Vert \overline{N}^{\varepsilon}\Vert _{H^{6}(\mathbf{Q})}^{6}\\
 & \leq & \const\underbrace{\sum_{m=1,2}
 \sum_{n=1,2}\left(\sum_{b^{\Delta}\in\mathcal{H}^{b^{\Delta S}}}
 \mathbf{E_{Q}}\left[\int_{0}^{T}\left|\frac{b_{t}^{\Delta}}
 {\sigma_{t}^{S}}\right|^{6m}dt\right]\right)^{\frac{1}{2n}}}_{<\infty}\varepsilon.
\end{eqnarray*}

\emph{2.} By (\ref{eq:stop_sigma}), Markov's inequality, 
(\ref{eq:sigma_delta-S<=00003D}), (\ref{eq:3-Eks-Ungl.}) for $q=2$, and Condition (\ref{eq:Bed.Dual_sigma}),
it holds that
\[
\mathbf{Q}(\rho^{\varepsilon,1}< T)\leq4\left\Vert \frac{c_{t}^{\Delta S,S}}{c_{t}^{S,S}}
\right\Vert _{S^{2}(\mathbf{Q})}^{2}\leq \const
\underbrace{\sum_{\sigma^{\Delta}\in\mathcal{H}^{\sigma^{\Delta S}}}
\left\Vert \frac{c_{t}^{\Delta,S}}{c_{t}^{S,S}}\right\Vert _{S^{2}(\mathbf{Q})}^{2}}_{<\infty}
\varepsilon^{4/3}.
\]

\emph{3.} Notice that
\[
\mathbf{Q}(\rho^{\varepsilon,2}< T)=\mathbf{Q}(\{ \rho^{\varepsilon,2}< T\} 
\cap\{ \rho^{\varepsilon,2}<\rho^{\varepsilon,1}\} )
+\mathbf{Q}(\rho^{\varepsilon,1}\leq\rho^{\varepsilon,2}< T),
\]
where
\[
\mathbf{Q}(\rho^{\varepsilon,1}\leq\rho^{\varepsilon,2}< T)\leq\mathbf{Q}(\rho^{\varepsilon,1}< T)
\overset{(\ref{eq:Ws(rho_sigma<T)})}{=}O\big(\varepsilon^{4/3}\big).
\]
From (\ref{eq:stop_Z}), Markov's inequality, (\ref{eq:S-Norm<=00003DH-Norm})
for $q=6$, (\ref{eq:Theta}, \ref{e:neudazu}, \ref{eq:Summen-Norm}), 
and Conditions (\ref{eq:Bed.Dual_sigma}, \ref{eq:Bed.Dual}), we deduce that
\begin{eqnarray*}
\lefteqn{\mathbf{Q}\left(\{ \rho^{\varepsilon,2}< T\} 
\cap\{ \rho^{\varepsilon,2}<\rho^{\varepsilon,1}\} \right)}\\
 & \leq & \mathbf{Q}\left(\exists t\leq T:|Z_{t}^{\varepsilon}-1|>\frac{1}{2}\right)\\
 & \leq & \mathbf{Q}\left(\exists t\leq T:|N_{t}^{\varepsilon}|>\ln\frac{3}{2}\right)\\
 & \leq & \const\mathbf{E_{Q}}\left[\sup_{t\in[0,T]}|N_{t}^{\varepsilon}|^{6}\right]\\
 & \leq & \const\left\Vert N^{\varepsilon}\right\Vert _{H^{6}(\mathbf{Q})}^{6}\\
 & \leq & \const\sum_{n=1,2} \Bigg(\sum_{b^{\Delta}\in\mathcal{H}^{b^{\Delta S}}}
 \underbrace{\mathbf{E_{Q}}\left[\int_{0}^{T}
 \left|\frac{b_{t}^{\Delta}}{\sigma_{t}^{S}}\right|^{12}dt\right]}_{<\infty}\Bigg)^{\frac1{2n}} \varepsilon^{2},
\end{eqnarray*}
which implies (\ref{eq:Ws(rho_Z<T)}).
\end{prf}

Now, let us pass to the convex duality theory. We denote by $\widetilde{\uti}$
the conjugate function of $\uti$, i.e.,
\begin{equation}
\widetilde{\uti}(y):=\sup_{x\in\mathbb{R}}(\uti(x)-xy),\quad y\geq0,\label{eq:konj.Fkt.}
\end{equation}
which satisfies $-\widetilde{\uti}^{\prime}=(\uti^{\prime})^{-1}$.
Since $\uti(x)=-e^{-px}$, we obtain
\begin{equation}
\widetilde{\uti}^{\prime}(y)=\frac{1}{p}\ln\frac{y}{p},\quad\widetilde{\uti}^{\prime\prime}(y)
=\frac{1}{py},\quad\widetilde{\uti}^{\prime\prime\prime}(y)=-\frac{1}{py^{2}}.\label{eq:konj._Exp.}
\end{equation}

\begin{lemma}\label{lem:Dual}
Conditions (\ref{eq:Bed.Primal}--\ref{eq:Bed.Dual}) imply
\begin{eqnarray}
\lefteqn{ \mathbf{E}\left[\widetilde{\uti}(yZ_{T}\overline{Z}_{T}^{\varepsilon})\right]+xy}\nonumber\\
 & = & \mathbf{E}[\uti(x+\varphi\mal S_{T})]-yp
 \mathbf{E_{Q}}\left[\Big((\Delta\varphi^{+})^{2}
 -\frac{3}{2}\Delta\varphi^{2}\Big)\mal[S,S]_{T}\right]+O(\varepsilon).
 \label{eq:loss}
\end{eqnarray}
\end{lemma}

\begin{prf}
By (\ref{eq:konj._Exp.}), Taylor expansion of $\widetilde{\uti}$
yields
\begin{eqnarray*}
\widetilde{\uti}(yZ_{T}\overline{Z}_{T}^{\varepsilon}) 
& = & \widetilde{\uti}(yZ_{T})\\
 &  & {}+\underbrace{\widetilde{\uti}^{\prime}(yZ_{T})yZ_{T}
 (\overline{Z}_{T}^{\varepsilon}-1)}_{=:G^{(14)}}+\underbrace{\frac{1}{2}
 \widetilde{\uti}^{\prime\prime}(yZ_{T})(yZ_{T})^{2}
 (\overline{Z}_{T}^{\varepsilon}-1)^{2}}_{=:G^{(15)}}\\
 &  &{} -\underbrace{\frac{1}{6p}\left(1+\theta(\overline{Z}_{T}^{\varepsilon}-1)\right)^{-1}
 yZ_{T}(\overline{Z}_{T}^{\varepsilon}-1)^{3}}_{=:G^{(16)}}
\end{eqnarray*}
for some random $\theta\in(0,1)$.
Due to the optimality of $\varphi$ and $yZ_{T}$ as well as by conjugate relations
(cf.\ \cite[Theorem 2.2]{schachermayer.99}), we conclude
\begin{equation}
\mathbf{E}[\widetilde{\uti}(yZ_{T})]=\mathbf{E}[\uti(x+\varphi\mal S_{T})]-xy,\label{eq:konj.Relation}
\end{equation}
\begin{equation}
-\widetilde{\uti}^{\prime}(yZ_{T})=x+\varphi\mal S_{T},\label{eq:-konj.'(Z)=00003DX}
\end{equation}
\begin{equation}
\widetilde{\uti}^{\prime\prime}(yZ_{T})(yZ_{T})^{2}=\frac{1}{p}yZ_{T}
,\label{eq:konj.''(Z)=00003D}
\end{equation}
and $\varphi\mal S$ is a $\mathbf{Q}$-martingale, i.e.,
\begin{equation}
\mathbf{E_{Q}}[x+\varphi\mal S_{T}]=x.\label{eq:opt.-M.-erzeugend}
\end{equation}
(\ref{eq:-konj.'(Z)=00003DX}, \ref{eq:opt.-M.-erzeugend}) yield
\[
\mathbf{E}\left[\widetilde{\Delta}_{1}\right]
=\underbrace{y\mathbf{E_{Q}}\left[x+\varphi\mal S_{T}\right]-xy}_{=0}
-y\mathbf{E_{Q}}\left[\overline{Z}_{T}^{\varepsilon}(\varphi\mal S_{T})\right]
\]
and
\begin{eqnarray*}
-\mathbf{E_{Q}}\left[\overline{Z}_{T}^{\varepsilon}(\varphi\mal S_{T})\right]
& = & -\mathbf{E_{Q}}\left[\overline{Z}_{T}^{\varepsilon}
(\overline{\varphi}^{\varepsilon}\mal\overline{S}_{T}^{\varepsilon})\right]
+\mathbf{E_{Q}}\left[\overline{\varphi}^{\varepsilon}\mal\overline{S}_{T}^{\varepsilon}-\varphi\mal S_{T}\right]\\
 &  &{} +\mathbf{E_{Q}}\left[(\overline{Z}_{T}^{\varepsilon}-1)
 (\overline{\varphi}^{\varepsilon}\mal\overline{S}_{T}^{\varepsilon}-\varphi\mal S_{T})\right].
\end{eqnarray*}
$\overline{Z}^{\varepsilon}(\overline{\varphi}^{\varepsilon}\mal\overline{S}^{\varepsilon})$
is a $\mathbf{Q}$-local martingale, cf.\ Remark~\ref{rem:Dual-EMM}.
(\ref{eq:stop_Z}, \ref{eq:liq<=00003D}, \ref{eq:stop_tao_2}) and Condition (\ref{eq:Bed.Primal})
yield
\[
\left\Vert \overline{Z}^{\varepsilon}(\overline{\varphi}^{\varepsilon}\mal\overline{S}^{\varepsilon})
\right\Vert _{S^{1}(\mathbf{Q})}\leq \const\left(\varepsilon^{-4/3}
+\mathbf{E_{Q}}\left[\sup_{t\in[0,T]}|\varphi_{t}S_{t}|
+\sup_{t\in[0,T]}|\beta_{t}S_{t}|\right]\right)<\infty,
\]
which implies that
$\overline{Z}^{\varepsilon}(\overline{\varphi}^{\varepsilon}\mal\overline{S}^{\varepsilon})$
is a uniformly integrable $\mathbf{Q}$-martingale and hence
\[
\mathbf{E_{Q}}[\overline{Z}_{T}^{\varepsilon}(\overline{\varphi}^{\varepsilon}\mal
\overline{S}_{T}^{\varepsilon})]=0.
\]
(\ref{eq:X-1_1}, \ref{eq:X-2_1}, \ref{eq:X-3_1}, \ref{eq:X-(Tao-T)_1 klein})
in conjunction with (\ref{eq:Ws(Tao<=00003DT)}, \ref{eq:Ws(rho_sigma<T)}, \ref{eq:Ws(rho_Z<T)})
and the argument in (\ref{eq:X-(Tao-T)_1 klein}) yield
\[
\mathbf{E_{Q}}\left[\overline{\varphi}^{\varepsilon}\mal\overline{S}_{T}^{\varepsilon}-\varphi\mal S_{T}\right]
=p\mathbf{E_{Q}}\left[\big(2\Delta\varphi^{2}
-(\Delta\varphi^{+})^{2}\big)\mal[S,S]_{\tau^{\varepsilon}\wedge\rho^{\varepsilon}}\right]
+O(\varepsilon).
\]
We have
\begin{eqnarray*}
\lefteqn{\mathbf{E_{Q}}\left[(\overline{Z}_{T}^{\varepsilon}-1)
(\overline{\varphi}^{\varepsilon}\mal\overline{S}_{T}^{\varepsilon}-\varphi\mal S_{T})\right]}\\
 & = & \underbrace{\mathbf{E_{Q}}\left[(\overline{Z}_{T}^{\varepsilon}-1)
 (\Delta\overline{\varphi}\mal S_{T})\right]}_{=:G^{(17)}}\\
 &  & {}+\underbrace{\mathbf{E_{Q}}\left[(\overline{Z}_{T}^{\varepsilon}-1)
 (\overline{\varphi}^{\varepsilon}\mal\Delta\overline{S}_{T}
 +\varphi\mal S_{\tau^{\varepsilon}\wedge\rho^{\varepsilon}}-\varphi\mal S_{T})\right]}_{=:G^{(18)}}.
\end{eqnarray*}
From (\ref{eq:X-1_2}), Cauchy-Schwarz' inequality and (\ref{eq:Delta-Z_2'}), we conclude
\begin{eqnarray*}
G^{(17)} & = & -p\mathbf{E_{Q}}\left[(\Delta\overline{\varphi}\mal S_{T})^{2}\right]
+\mathbf{E_{Q}}\left[(\overline{Z}_{T}^{\varepsilon}-1
+p\Delta\overline{\varphi}\mal S_{T})(\Delta\overline{\varphi}\mal S_{T})\right]\\
 & = & -p\mathbf{E_{Q}}\left[\Delta\overline{\varphi}^{2}\mal[S,S]_{T}\right]+O(\varepsilon).
\end{eqnarray*}
By Cauchy-Schwarz' inequality, (\ref{eq:stop_Z}, \ref{eq:Delta-Z_2}), 
using (\ref{eq:X-2_2-6}, \ref{eq:X-3_2-6}, \ref{eq:3-Eks-Ungl.}) for $q=2$, and combining the argument in
(\ref{eq:X-(Tao-T)_1 klein}) with (\ref{eq:Ws(rho_sigma<T)}, \ref{eq:Ws(rho_Z<T)}), it follows that
\begin{eqnarray*}
|G^{(18)}| & \leq & \const\sqrt{\mathbf{E_{Q}}\left[(\overline{Z}_{T}^{\varepsilon}-1)^{2}\right]}
\sqrt{\mathbf{E_{Q}}\left[(\overline{\varphi}^{\varepsilon}\mal\Delta\overline{S}_{T})^{2}\right]}\\
 &  & {}+\const\mathbf{E_{Q}}\left[\mathbf{1}_{\{ \tau^{\varepsilon}\wedge\rho^{\varepsilon}< T\} }
 |\varphi\mal S_{\tau^{\varepsilon}\wedge\rho^{\varepsilon}}-\varphi\mal S_{T}|\right]\\
 & = & O(\varepsilon).
\end{eqnarray*}
Together, we obtain
\begin{equation}
\mathbf{E}\left[G^{(14)}\right]
=yp\mathbf{E_{Q}}\left[\big(\Delta\varphi^{2}
-(\Delta\varphi^{+})^{2}\big)\mal[S,S]_{\tau^{\varepsilon}\wedge\rho^{\varepsilon}}\right]
+O(\varepsilon).\label{eq:Delta-Z_1}
\end{equation}
(\ref{eq:konj.''(Z)=00003D}) and (\ref{eq:Delta-Z_2}) yield
\begin{eqnarray}
\mathbf{E}\left[G^{(15)}\right] 
& = & \frac{y}{2p}\mathbf{E_{Q}}\left[(\overline{Z}_{T}^{\varepsilon}-1)^{2}\right]\nonumber \\
 & = & \frac{yp}{2}\mathbf{E_{Q}}\left[\Delta\overline{\varphi}^{2}\mal[S,S]_{T}\right]+O(\varepsilon).
 \label{eq:Delta-Z_2 eingesetzt}
\end{eqnarray}
By (\ref{eq:stop_Z}) and (\ref{eq:Delta-Z_3}), we have
\begin{equation}
\left|\mathbf{E}\left[G^{(16)}\right]\right|
\leq \const\mathbf{E_{Q}}\left[|\overline{Z}_{T}^{\varepsilon}-1|^{3}\right]
=O(\varepsilon).\label{eq:Rest<Delta-Z_3}
\end{equation}
From (\ref{eq:konj.Relation}, \ref{eq:Delta-Z_1}, \ref{eq:Delta-Z_2 eingesetzt}, \ref{eq:Rest<Delta-Z_3}) 
we obtain
\begin{eqnarray*}
\lefteqn{\mathbf{E}\left[\widetilde{\uti}(yZ_{T}\overline{Z}_{T}^{\varepsilon})\right]+xy}\\
 & = & \mathbf{E}\left[\uti(x+\varphi\mal S_{T})\right]
 -yp\mathbf{E_{Q}}\left[\Big((\Delta\varphi^{+})^{2}
 -\frac{3}{2}\Delta\varphi^{2}\Big)\mal[S,S]_{\tau^{\varepsilon}\wedge\rho^{\varepsilon}}\right]
 +O(\varepsilon).
\end{eqnarray*}
Combining this with (\ref{eq:Ws(rho_sigma<T)}, \ref{eq:Ws(rho_Z<T)}) and the argument in
 (\ref{eq:Verlust(Tao-T) klein}), the assertion follows.
\end{prf}

\subsection{Optimality}
Having approximated both the primal and dual value of the optimization problem, we are now
able to prove the leading-order optimality of the candidate strategy $\varphi^{\varepsilon}$.

\begin{lemma}\label{thm:Opt}
Under Assumption \ref{ass:Primal} we have
\[
\sup_{\psi\in\mathcal{A}^{\varepsilon}(x^{B},x^{S})}
\mathbf{E}\left[\uti(X_{T}^{\psi,\varepsilon})\right]=
\mathbf{E}\left[\uti\big(X_{\tau^{\varepsilon}}^{\varphi^{\varepsilon},\varepsilon}\big)\right]
+O(\varepsilon).
\]
\end{lemma}

\begin{prf}
Take an arbitrary admissible trading strategy $\psi\in\mathcal{A}^{\varepsilon}(x^{B},x^{S})$
and let $\overline{S}^{\varepsilon}$, $\overline{Z}^{\varepsilon}$ be as in (\ref{eq:Def.stop_phi, S, Z}). 
Since $\overline{S}^{\varepsilon}$ has values in $[(1-\varepsilon)S,(1+\varepsilon)S]$, we have
\begin{equation}
x+\psi\mal\overline{S}^{\varepsilon}\geq X^{\psi,\varepsilon}\geq-K\quad \textrm{a.s.}\label{eq:Schatten>liq}
\end{equation}
for some $K\in\mathbb{R}_{+}$.
Recalling Remark~\ref{rem:Dual-EMM}, 
$\overline{Z}^{\varepsilon}(x+\psi\mal\overline{S}^{\varepsilon})$
is a $\mathbf{Q}$-local martingale and hence a $\mathbf{Q}$-supermartingale
by (\ref{eq:Schatten>liq}). Therefore
\begin{equation}
\mathbf{E_{Q}}\left[\overline{Z}_{T}^{\varepsilon}
(x+\psi\mal\overline{S}_{T}^{\varepsilon})\right]\leq x.\label{eq:super}
\end{equation}
Together, we conclude
\begin{eqnarray*}
\mathbf{E}\left[\uti(X_{T}^{\psi,\varepsilon})\right] & \overset{(\ref{eq:Schatten>liq})}{\leq} 
& \mathbf{E}\left[\uti(x+\psi\mal\overline{S}_{T}^{\varepsilon})\right]\\
& \overset{(\ref{eq:konj.Fkt.})}{\leq} & \mathbf{E}\left[\widetilde{\uti}(yZ_{T}
\overline{Z}_{T}^{\varepsilon})\right]+y\mathbf{E_{Q}}\left[\overline{Z}_{T}^{\varepsilon}
(x+\psi\mal\overline{S}_{T}^{\varepsilon})\right]\\
& \overset{(\ref{eq:super})}{\leq} & \mathbf{E}\left[\widetilde{\uti}(yZ_{T}
\overline{Z}_{T}^{\varepsilon})\right]+xy\\
& \overset{\textrm{Lemma~\ref{lem:Dual}}}{=} & 
\mathbf{E}\left[\uti(x+\varphi\mal S_{T})\right]\\
&  & {}-yp\mathbf{E_{Q}}\left[\Big((\Delta\varphi^{+})^{2}-\frac{3}{2}
\Delta\varphi^{2}\Big)\mal[S,S]_{T}\right]+O(\varepsilon)\\
& \overset{\textrm{Lemma~\ref{lem:Primal}}}{=} & 
\mathbf{E}\left[\uti(X_{\tau^{\varepsilon}}^{\varphi^{\varepsilon},\varepsilon})\right]
+O(\varepsilon).
\end{eqnarray*}
Since
$\varphi^{\varepsilon}\mathbf{1}_{\auf0,\tau^{\varepsilon}\zu}
\in\mathcal{A}^{\varepsilon}(x^{B},x^{S})$ by definition, 
this proves the assertion.
\end{prf}

\subsection{Certainty equivalent loss}
In this section we  express the minimal  loss of utility caused
by transaction costs in terms of $\Delta\varphi^{+}$ rather than both $\Delta\varphi^{+}$ and
$\Delta\varphi$, cf.\ (\ref{e:imlemma}, \ref{eq:loss}).
Throughout this section, we suppose that Assumption \ref{ass:Cont.} holds.

Set
\[
q:=\frac{\Delta\varphi}{\Delta\varphi^{+}}.
\]
Then $q$ is a semimartingale reflected to stay between $\pm1$.
By Itô's formula, its dynamics are characterized by
\[
dq_{t}=b_{t}^{q}dt+dM_{t}^{q,\mathbf{Q}}+dA_{t}^{+}-dA_{t}^{-},
\]
where $M^{q,\mathbf{Q}}$ is a continuous $\mathbf{Q}$-local martingale
starting in $0$, processes $A^{+}$ and $A^{-}$ are increasing processes 
which  grow only on $\{ q=-1\} $ and $\{ q=1\} $, respectively,
and
\begin{equation}
b_{t}^{q}=-\Big(\frac{2}{\varepsilon}\Big)^{1/3}\Big(\frac{b_{t}^{\varphi}}
{\beta_{t}}+c_{t}^{\varphi,{1}/{\beta}}\Big)-\frac{q_{t}}{\beta_{t}}
b_{t}^{\beta}+\frac{q_{t}}{\beta_{t}^{2}}c_{t}^{\beta,\beta},\label{eq:b_q}
\end{equation}
\begin{equation}
c_{t}^{q,q}:=\frac{d[q,q]_{t}}{dt}
=\Big(\frac{2}{\varepsilon}\Big)^{2/3}
\frac{c_{t}^{\varphi,\varphi}}{\beta_{t}^{2}}
+\left(\frac{16}{\varepsilon}\right)^{1/3}\frac{q_{t}}{\beta_{t}^{2}}c_{t}^{\beta,\varphi}
+\frac{q_{t}^{2}}{\beta_{t}^{2}}c_{t}^{\beta,\beta}.\label{eq:c_q,q}
\end{equation}
Define a stopping time $\sigma^{\varepsilon}:=\inf\{ t\in[0,T]:|b_{t}^{q}|>
\frac{1}{\varepsilon}\textrm{ or } c_{t}^{q,q}<\varepsilon\} \wedge T$
and let $\overline{q}:=q^{\sigma^{\varepsilon}}$.

\subsubsection{Time change}\label{subsec:timechange}
Fix $t\in[0,T)$. Consider the time change $(t(\vartheta))_{\vartheta\in\mathbb R_+}$
defined by
\[
t(\vartheta):=\inf\left\{ s\in[t,T]:[q,q]_{s}-[q,q]_{t}>\vartheta\right\} \wedge\sigma^{\varepsilon}.
\]
Set
\[
\overline{\vartheta}:=[\overline q,\overline q]_{T}
-[\overline q,\overline q]_{t}
=\int_t^Tc^{\overline q,\overline q}_sds
\]
and $\widetilde{q}_{\vartheta}:=\overline{q}_{t(\vartheta)}$ for
$\vartheta\in\mathbb{R}_{+}$. 

Fix $\omega\in\Omega$. For $\varepsilon>0$
small enough we have that $c^{q,q}(\omega)$ exceeds $\varepsilon$ on $[0,T]$. 
Therefore, the mapping $\vartheta\mapsto t(\vartheta)$
is continuously differentiable on the interval $(0,\overline\vartheta)$ 
with derivative $(c_{t(\vartheta)}^{q,q})^{-1}$.

\begin{lemma}
Recall that Assumption \ref{ass:Cont.} is supposed to hold.
Setting
\[
 \vartheta^{\varepsilon}:=
 \int_{t}^{(t+\varepsilon^{1/3})\wedge T}c_{s}^{\overline q,\overline q}ds,
\]
we have
\begin{equation}
\lim_{\varepsilon\downarrow0}\left|\frac{1}{\varepsilon^{1/3}}
\int_{t}^{(t+\varepsilon^{1/3})\wedge T}q_{s}^{2}
\Big(\frac{\Delta\varphi_{s}^{+}}{\varepsilon^{1/3}}\Big)^{2}c_{s}^{S,S}ds
-\frac{1}{\vartheta^{\varepsilon}}\int_{0}^{\vartheta^{\varepsilon}}
\widetilde{q}_{\vartheta}^{2}d\vartheta\Big(\frac{\Delta\varphi_{t}^{+}}
{\varepsilon^{1/3}}\Big)^{2}c_{t}^{S,S}\right|=0\quad \textrm{a.s.}\label{eq:f.s.-Konv.}
\end{equation}
For any $\omega\in\Omega$
there exists some $\varepsilon_{0}(\omega)$, $\underline{K}(\omega),\overline{K}(\omega)>0$
such that
\begin{equation}
\underline{K}(\omega)\varepsilon^{-1/3}
\leq\vartheta^{\varepsilon}(\omega)\leq\overline{K}(\omega)\varepsilon^{-1/3}\label{eq:Horizont}
\end{equation}
holds for any $\varepsilon\leq\varepsilon_{0}(\omega)$.
\end{lemma}

\begin{prf}
Fix $\omega\in\Omega$
and consider events
\[
A^{\varepsilon,b}:=\left\{ \exists t\in[0,T]:|b_{t}^{q}|>\frac{1}{\varepsilon}\right\} ,
\quad A^{\varepsilon,c}:=\left\{ \exists t\in[0,T]:c_{t}^{q,q}<\varepsilon\right\} .
\]
Since all processes in (\ref{eq:b_q}) are assumed to have continuous or at least bounded
paths, there exists $C(\omega)<\infty$ 
such that
\[
\sup_{t\in[0,T]}|b_{t}^{q}|(\omega)\leq C(\omega)\varepsilon^{-1/3},
\]
whence
$\omega\notin A^{\varepsilon,b}$
for any $\varepsilon$ that is small enough.
Similarly, there exists $c(\omega)>0$ 
such that
\[
\min_{t\in[0,T]}\frac{c_{t}^{\varphi,\varphi}}{\beta_{t}^{2}}(\omega)>c(\omega)
\]
and hence $\omega\notin A^{\varepsilon,c}$ for any $\varepsilon$ that is small enough.
Therefore $\sigma^{\varepsilon}(\omega)=T$ for $\varepsilon$
small enough, which implies that
\begin{equation}
\lim_{\varepsilon\downarrow0}\left|\frac{1}{\varepsilon^{1/3}}
\int_{t}^{(t+\varepsilon^{1/3})\wedge T}(q_{s}^{2}
-\overline{q}_{s}^{2})\Big(\frac{\Delta\varphi_{s}^{+}}
{\varepsilon^{1/3}}\Big)^{2}c_{s}^{S,S}ds\right|(\omega)=0.\label{eq:stopp_q}
\end{equation}
By continuity of the mapping $s\mapsto(\frac{\Delta\varphi_{s}^{+}}
{\varepsilon^{1/3}})^{2}c_{s}^{S,S}(\omega)$
at $t$ and using the mean value theorem, we have
\begin{equation}
\lim_{\varepsilon\downarrow0}\left|\frac{1}{\varepsilon^{1/3}}
\int_{t}^{(t+\varepsilon^{1/3})\wedge T}\overline{q}_{s}^{2}
\left(\Big(\frac{\Delta\varphi_{s}^{+}}{\varepsilon^{1/3}}\Big)^{2}c_{s}^{S,S}
-\Big(\frac{\Delta\varphi_{t}^{+}}{\varepsilon^{1/3}}\Big)^{2}c_{t}^{S,S}\right)ds\right|(\omega)=0.
\label{eq:MWS_t}
\end{equation}
Applying the mean value theorem to the mapping $t\mapsto t(\vartheta)$, we get
\begin{eqnarray}
\varepsilon^{1/3}
&=&\left(t(\vartheta^\varepsilon)-t(0)\right)(\omega)\nonumber\\
&=&\left((c_{t(\xi)}^{\overline q,\overline q})^{-1}\vartheta^{\varepsilon}\right)(\omega)
\textrm{ for some }\xi\in[0,\vartheta^{\varepsilon}(\omega)]\label{eq:MWS_Horizont}
\end{eqnarray}
 for $\varepsilon$ small enough
and
\[
\lim_{\varepsilon\downarrow0}\left|\frac{1}{\varepsilon^{1/3}}
\int_{0}^{\vartheta^{\varepsilon}}\widetilde{q}_{\vartheta}^{2}
\left(\big(c_{t(\vartheta)}^{\overline q,\overline q}\big)^{-1}
-\big(c_{t(\xi)}^{\overline q,\overline q}\big)^{-1}\right)d\vartheta\right|(\omega)=0.
\]
Change of variables yields
\begin{equation}
\lim_{\varepsilon\downarrow0}\left|\frac{1}{\varepsilon^{1/3}}
\int_{t}^{(t+\varepsilon^{1/3})\wedge T}\overline{q}_{s}^{2}ds
-\frac{1}{\vartheta^{\varepsilon}}\int_{0}^{\vartheta^{\varepsilon}}
\widetilde{q}_{\vartheta}^{2}d\vartheta\right|(\omega)=0.\label{eq:MWS_theta}
\end{equation}

Combining (\ref{eq:stopp_q}, \ref{eq:MWS_t}, \ref{eq:MWS_theta})
yields (\ref{eq:f.s.-Konv.}). Moreover, (\ref{eq:Horizont}) follows
from (\ref{eq:MWS_Horizont}, \ref{eq:c_q,q}) and continuity of
the coefficients in (\ref{eq:c_q,q}).
\end{prf}

\subsubsection{Change of measure}
We use the same notation as in Section~\ref{subsec:timechange}. 
From the Dambis-Dubins-Schwarz theorem (cf.\ \cite[Theorems V.1.6, V.1.7]{revuz.yor.99}),
there exists an enlargement 
$(\widetilde{\Omega},\widetilde{\mathcal{F}},
(\widetilde{\mathcal{F}}_\vartheta)_{\vartheta\in\mathbb{R}_{+}},\widetilde{\mathbf{Q}})$
of the filtered space
$(\Omega,\mathcal{F},(\mathcal{F}_{t(\vartheta)})_{\vartheta\in\mathbb{R}_{+}},\mathbf{Q})$
and a standard Brownian motion $\widetilde{W}^{\mathbf{Q}}$ on that space
such that 
$ \widetilde{W}_{\vartheta}^{\mathbf{Q}}=M_{t(\vartheta)}^{q,\mathbf{Q}}-M_{t(0)}^{q,\mathbf{Q}}$
for $\vartheta<\overline{\vartheta}$.

Since the process
$(\widetilde{b}_{\vartheta})_{\vartheta\in\mathbb{R}_{+}}$
defined by 
\[\widetilde{b}_{\vartheta}:=
 \frac{b_{t(\vartheta)}^{\overline{q}}}{c_{t(\vartheta)}^{\overline{q},\overline{q}}}
 \mathbf{1}_{\auf0,\overline\vartheta\zu}(\vartheta)
\]
is bounded,
\[
\frac{d\mathbf{Q}^{\varepsilon}}{d\widetilde{\mathbf{Q}}}
=\exp\left(-\int_{0}^{\vartheta^{\varepsilon}}\widetilde{b}_{\vartheta}
d\widetilde{W}_{\vartheta}^{\mathbf{Q}}-\frac{1}{2}\int_{0}^{\vartheta^{\varepsilon}}
\widetilde{b}_{\vartheta}^{2}d\vartheta\right)
\]
defines a probability measure on $(\widetilde{\Omega},\widetilde{\mathcal{F}},
(\widetilde{\mathcal{F}}_\vartheta)_{\vartheta\in\mathbb{R}_{+}})$
whose Hellinger process $h(\frac12,\mathbf{Q}^{\varepsilon},\mathbf{Q})$ is given by
\[
 \textstyle h(\frac12,\mathbf{Q}^{\varepsilon},\widetilde{\mathbf{Q}})_\vartheta=
 \int_{0}^{\vartheta\wedge\vartheta^{\varepsilon}}
\widetilde{b}_{\zeta}^{2}d\zeta,\quad \vartheta\in\mathbb R_+
\]
(cf.\ \cite[Theorem IV.1.33]{js.87}) and
such that
\[
\widetilde{W}^{\mathbf{Q}^{\varepsilon}}:=\widetilde{W}^{\mathbf{Q}}+\int_{0}^{\cdot}
\mathbf{1}_{\auf0,\vartheta^{\varepsilon}\zu}(\vartheta)\widetilde{b}_{\vartheta}d\vartheta
\]
is a $\mathbf{Q}^{\varepsilon}$-standard Brownian motion. 
In view of (\ref{eq:b_q}, \ref{eq:c_q,q}, \ref{eq:Horizont}),
we have
\[
\lim_{\varepsilon\downarrow0}\int_{0}^\infty\mathbf{1}_{\auf0,\vartheta^{\varepsilon}\zu}(\vartheta)
\widetilde{b}_{\vartheta}^{2}d\vartheta=0\quad\textrm{a.s.}
\]
By \cite[Theorem V.4.31 and Lemma V.4.3]{js.87} this implies
\begin{equation}
\lim_{\varepsilon\downarrow0}\sup_{A\in\widetilde{\mathcal F}_\infty}
|\mathbf{Q}^{\varepsilon}(A)-\widetilde{\mathbf{Q}}(A)|=0.\label{eq:meas.}
\end{equation}

\begin{lemma}
On the probability space
$(\Omega,\mathcal F,\mathbf Q)$, we have
\begin{equation}
\left|\frac{1}{\vartheta^{\varepsilon}}
\int_{0}^{\vartheta^{\varepsilon}}\widetilde{q}_{\vartheta}^{2}d\vartheta-\frac13\right|
\stackrel{\varepsilon\downarrow0}\longrightarrow0
\quad \textrm{ in probability.}\label{eq:LLN}
\end{equation}
\end{lemma}

\begin{prf}
Let process $Y$ starting at $\widetilde{q}_{0}$ be the unique solution
to the Skorohod SDE
\[
dY_{\vartheta}=d\widetilde{W}_{\vartheta}^{\mathbf{Q}^{\varepsilon}}
\]
with reflection at $\pm1$ (cf.\ e.g.\  \cite[Theorem 3.3]{slominski.wojciechowski.13}
for existence and uniqueness).
Observe that $Y$ coincides with $\widetilde{q}$ on $\auf0,\vartheta^{\varepsilon}\zu$.
Indeed, according to \cite[10.18]{jacod.79}, we have
\[
d\widetilde{q}_{\vartheta}=\mathbf{1}_{\auf0,\overline{\vartheta}\zu}(\vartheta)
d\widetilde{W}_{\vartheta}^{\mathbf{Q}}
+\widetilde{b}_{\vartheta}d\vartheta
+dA^{+}_{t(\vartheta)}-dA^{-}_{t(\vartheta)},
\]
i.e., $\widetilde{q}$ solves the Skorohod SDE on $\auf0,\overline{\vartheta}\zu$,
which yields $Y=\widetilde{q}$ on $\auf0,\vartheta^{\varepsilon}\zu$
by uniqueness of the solution to the stopped Skorohod SDE. Note that
standard Brownian motion reflected at $\pm1$ is a Markov process
with uniform stationary distribution, cf.\ e.g.\ \cite[Appendix 1.5]{borodin.salminen.02}.

Let $\delta>0$.
Due to \cite[Theorem]{katz.thomasian.60} and in view of (\ref{eq:Horizont}), there exist
two constants $C<\infty$ and $\varsigma<1$ such that for all $\varepsilon_{0}\in(0,1),
\varepsilon\leq\varepsilon_{0},\vartheta\in[0,1]$ we have
\[
 \mathbf{Q}^{\varepsilon}\left(A_{\vartheta}^{\varepsilon}\right)<C\varsigma^{\frac{1}{\varepsilon_{0}}}
\]
for 
\[
 A_{\vartheta}^{\varepsilon}:=\left\{\left|\frac{1}{\lfloor \vartheta^{\varepsilon}\rfloor }
\sum_{i=0}^{\lfloor \vartheta^{\varepsilon}\rfloor -1}(\widetilde{q}_{\vartheta+i})^{2}
-\frac13\right|>\delta\right\}.
\]
In combination with (\ref{eq:meas.})
and interpreting $A_{\vartheta}^{\varepsilon}$ naturally as a subset of $\Omega$, we obtain
\[
\lim_{\varepsilon\downarrow0}\mathbf{Q}\left(A_{\vartheta}^{\varepsilon}\right)
=\lim_{\varepsilon\downarrow0}\widetilde{\mathbf{Q}}\left(A_{\vartheta}^{\varepsilon}\right)=0.
\]
Fubini's theorem and dominated convergence yield
\[
 \lim_{\varepsilon\downarrow0}\mathbf{E_{Q}}\left[\int_{0}^{1}\mathbf{1}_{A_{\vartheta}^{\varepsilon}}
 d\vartheta\right]
=\lim_{\varepsilon\downarrow0}\int_{0}^{1}\mathbf{Q}(A_{\vartheta}^{\varepsilon})d\vartheta
=\int_{0}^{1}
\lim_{\varepsilon\downarrow0}\mathbf{Q}(A_{\vartheta}^{\varepsilon})d\vartheta=0.
\]
Again by dominated convergence we obtain
\[
\mathbf{E}_\mathbf{Q}\left[\left|\int_{0}^{1}
 \bigg(\frac{1}{\lfloor \vartheta^{\varepsilon}\rfloor }
 \sum_{i=0}^{\lfloor \vartheta^{\varepsilon}\rfloor -1}
 (\widetilde{q}_{\vartheta+i})^{2}\bigg)d\vartheta-\frac13\right|\right]
 \leq
 \mathbf{E}_\mathbf{Q}\left[\int_{0}^{1}
\bigg| \frac{1}{\lfloor \vartheta^{\varepsilon}\rfloor }
 \sum_{i=0}^{\lfloor \vartheta^{\varepsilon}\rfloor-1 }
 (\widetilde{q}_{\vartheta+i})^{2}-\frac13\bigg|d\vartheta\right]
 \to0
\]
and hence
\[
 \left|\int_{0}^{1}
 \bigg(\frac{1}{\lfloor \vartheta^{\varepsilon}\rfloor }
 \sum_{i=0}^{\lfloor \vartheta^{\varepsilon}\rfloor -1}
 (\widetilde{q}_{\vartheta+i})^{2}\bigg)d\vartheta-\frac13\right|
 \to 0
\mbox{ in probability}\]
for $\varepsilon\downarrow0$.
Since
\[
\frac{1}{\vartheta^{\varepsilon}}
 \int_{0}^{\vartheta^{\varepsilon}}(\widetilde{q}_{\vartheta})^{2}d\vartheta
 -\int_{0}^{1}
 \bigg(\frac{1}{\lfloor \vartheta^{\varepsilon}\rfloor }
 \sum_{i=0}^{\lfloor \vartheta^{\varepsilon}\rfloor -1}
 (\widetilde{q}_{\vartheta+i})^{2}\bigg)d\vartheta\to0\quad\textrm{a.s.}
\]
as $\varepsilon\downarrow0$, the assertion follows.
\end{prf}

\subsubsection{Asymptotics}
Gathering the previous considerations, we are now able to complete
our arguments concerning the welfare impact of small transaction costs.

\begin{lemma}\label{thm:Nutzenverlust}
Under Assumptions \ref{ass:Primal}, \ref{ass:Cont.} we have
\[
\mathbf{E_{Q}}\left[(\Delta\varphi)^{2}\mal[S,S]_{T}\right]
=\frac13\mathbf{E_{Q}}\left[(\Delta\varphi^{+})^{2}\mal[S,S]_{T}\right]
+o\big(\varepsilon^{2/3}\big).
\]
\end{lemma}

\begin{prf}
{\em Step 1:}
Let $\delta>0$ be arbitrary. 
Fix $t\in[0,T)$ and define $\vartheta^\varepsilon, \widetilde{q}$ as in Section \ref{subsec:timechange}.
Let
\[
\Delta X_{t}^{\varepsilon}:=\frac{1}{\varepsilon^{1/3}}
\int_{t}^{(t+\varepsilon^{1/3})\wedge T}q_{s}^{2}
\Big(\frac{\Delta\varphi_{s}^{+}}{\varepsilon^{1/3}}\Big)^{2}c_{s}^{S,S}ds
\]
and
\[
 A^\varepsilon_t:=\left\{\left|\Delta X_{t}^{\varepsilon}-\frac13
\Big(\frac{\Delta\varphi_{t}^{+}}{\varepsilon^{1/3}}\Big)^{2}c_{t}^{S,S}\right|>\delta\right\}.
\]
 By (\ref{eq:f.s.-Konv.}) and (\ref{eq:LLN}), we have
\begin{eqnarray*}
0 & \leq & \lim_{\varepsilon\downarrow0}\mathbf{Q}\left(A^\varepsilon_t\right)\\
 & \leq & \lim_{\varepsilon\downarrow0}\mathbf{Q}\left(\left|\Delta X_{t}^{\varepsilon}
 -\frac{1}{\vartheta^{\varepsilon}}\int_{0}^{\vartheta^{\varepsilon}}
 \widetilde{q}_{\vartheta}^{2}d\vartheta\Big(\frac{\Delta\varphi_{t}^{+}}
 {\varepsilon^{1/3}}\Big)^{2}c_{t}^{S,S}\right|>\frac{\delta}{2}\right)\\
 &  & {}+\lim_{\varepsilon\downarrow0}\mathbf{Q}\left(\left|\frac{1}{\vartheta^{\varepsilon}}
 \int_{0}^{\vartheta^{\varepsilon}}\widetilde{q}_{\vartheta}^{2}d\vartheta-
 \frac13\right|2^{-1/3}\beta_tc^{S,S}_t>\frac{\delta}{2}\right)\\
 & = & 0.
\end{eqnarray*}
 
{\em Step 2:}
Fubini's theorem, dominated convergence, and Step 1 yield
\begin{equation}\label{e:produktkonvergenz}
 \lim_{\varepsilon\downarrow0}\mathbf{E_{Q}}\left[\int_{0}^{T}\mathbf{1}_{A_{t}^{\varepsilon}}dt\right]
=\lim_{\varepsilon\downarrow0}\int_{0}^{T}\mathbf{Q}(A_{t}^{\varepsilon})dt=\int_{0}^{T}
\lim_{\varepsilon\downarrow0}\mathbf{Q}(A_{t}^{\varepsilon})dt=0.
\end{equation}
Observe that for any $t\in[0,T]$,
\[
\Delta X_{t}^{\varepsilon}\leq\frac{1}{\varepsilon^{1/3}}
\int_{t}^{(t+\varepsilon^{1/3})\wedge T}
\Big(\frac{\Delta\varphi_{s}^{+}}{\varepsilon^{1/3}}\Big)^{2}c_{s}^{S,S}ds
=:\Delta X_{t}^{\varepsilon+}.
\]
Using Fubini's theorem, we conclude
\begin{equation}\label{e:nocheiner}
\int_{\varepsilon^{1/3}}^T\ \Big(\frac{\Delta\varphi_s}{\varepsilon^{1/3}}\Big)^{2}c^{S,S}_sds
\leq\int_{0}^{T}\Delta X_{t}^{\varepsilon}dt
\leq\int_0^T\Big(\frac{\Delta\varphi_s}{\varepsilon^{1/3}}\Big)^{2}c^{S,S}_sds
\end{equation}
and
\begin{equation}\label{e:majorante}
\int_{0}^{T}\Delta X_{t}^{\varepsilon+}dt
\leq\int_0^T\Big(\frac{\Delta\varphi^{+}_s}{\varepsilon^{1/3}}\Big)^{2}c^{S,S}_sds.
\end{equation}
So by Condition (\ref{eq:Bed.Primal}),
\[
\mathbf{E_{Q}}\left[\int_{0}^{T}
\sup_{\varepsilon\in(0,1)}|\Delta X_{t}^{\varepsilon}|dt\right]
\leq\mathbf{E_{Q}}\left[\Big(\frac{\Delta\varphi^{+}}{\varepsilon^{1/3}}\Big)^{2}\mal[S,S]_{T}\right]<\infty.
\]
The assertion follows from (\ref{e:produktkonvergenz}, \ref{e:nocheiner}, \ref{e:majorante}) and dominated
convergence.
\end{prf}

\begin{cor}\label{cor:Nutzenverlust}
Under Assumptions~\ref{ass:Primal}, \ref{ass:Cont.} we have
\[
\sup_{\psi\in\mathcal{A}^{\varepsilon}(x^{B},x^{S})}\mathbf{E}\left[\uti(X_{T}^{\psi,\varepsilon})\right]
=\mathbf{E}\left[\uti(x+\varphi\mal S_{T})\right]
-\frac{yp}{2}\mathbf{E_{Q}}\left[(\Delta\varphi^{+})^{2}\mal[S,S]_{T}\right]+o\big(\varepsilon^{2/3}\big)
\]
and hence
\[
\sup_{\psi\in\mathcal{A}^{\varepsilon}(x^{B},x^{S})}\mathbf{CE}(X_{T}^{\psi,\varepsilon})
=\mathbf{CE}(x+\varphi\mal S_{T})
-\frac{p}{2}\mathbf{E_{Q}}\left[(\Delta\varphi^{+})^{2}\mal[S,S]_{T}\right]+o\big(\varepsilon^{2/3}\big).
\]
\end{cor}

\begin{prf}
The assertion follows from Lemmas~\ref{lem:Primal}, \ref{thm:Nutzenverlust}, \ref{thm:Opt}
and Taylor expansion of $y\mapsto -\frac1p\ln(-y)$ at
$\mathbf{E}\left[\uti(x+\varphi\mal S_{T})\right]$.
\end{prf}

\begin{appendix}\section{Appendix}
As an auxiliary result, we determine the explicit solution to the frictionless
optimization problem related to the stochastic volatility model in Section~\ref{subsec:Vol-Modell}.
We proceed analogously as in \cite[Theorem 3.1]{kallsen.muhlekarbe.09c}, which deals with power utility.

\begin{thm}\label{thm:Volphi}
For the stochastic volatility model characterized by (\ref{eq:verallg. BS})
with bounded ${b(Y)/\sigma(Y)}$,
the frictionless optimizer $\varphi$ satisfies
\begin{equation}\label{eq:opt}
\varphi_{t}S_{t}=\frac{b(Y_{t})}{p\sigma(Y_{t})^{2}}\quad\textrm{for all }t\in[0,T]. 
\end{equation}
The MEMM $\mathbf{Q}$ has density process
\[
\mathbf{E}\left[\left.\frac{d\mathbf{Q}}{d\mathbf{P}}\right|\mathcal{F}_{t}\right]
=\frac{\widetilde{Z}_{t}}{\widetilde{Z}_{0}}\exp\left(-\int_{0}^{t}\frac{b(Y_{s})}{\sigma(Y_{s})}dW_{s}
-\frac{1}{2}\int_{0}^{t}\Big(\frac{b(Y_{s})}{\sigma(Y_{s})}\Big)^{2}ds\right),
\quad t\in[0,T],
\]
where the process $\widetilde{Z}$ is defined as in (\ref{eq:verallg. BS-Z}).
\end{thm}

\begin{prf}
\emph{Step 1:}
Define filtration $\mathbf{G}=(\mathcal{G}_{t})_{t\in[0,T]}$
by
\begin{equation}
\mathcal{G}_{t}:=\bigcap_{s>t}\sigma\left(\mathcal{F}_{s}\cup
\sigma\big((Y_{r})_{r\in[0,T]}\big)\right),\quad t\in[0,T]\label{eq:Filt.G}
\end{equation}
and let
\[
\varphi:=\frac{b(Y)}{p\sigma(Y)^{2}S}
\]
in line with (\ref{eq:opt}).
Moreover, set
\begin{equation}
 \overline{Z}_{t}:=\exp\left(-\int_{0}^{t}\frac{b(Y_{s})}{\sigma(Y_{s})}dW_{s}
-\frac{1}{2}\int_{0}^{t}\Big(\frac{b(Y_{s})}{\sigma(Y_{s})}\Big)^{2}ds\right),
\quad t\in[0,T].\label{e:Zquer}
\end{equation}
By definition of $\mathbf{G}$, random variable $\widetilde{Z}_{T}$ is $\mathcal{G}_{0}$-measurable.
Since $Y$ is independent of $W$, it follows from \cite[Theorem 15.5]{bauer.02} 
that $W$ is a standard Brownian motion with respect to $\mathbf{G}$ as well.
Due to boundedness of ${b(Y)}/{\sigma(Y)}$,
 the local martingale $\overline{Z}$ satisfies Novikov's
condition, whence it is a martingale relative to both $\mathbf{F}$
and $\mathbf{G}$. Therefore, we deduce that
\begin{eqnarray}
Z_{t}^{\mathbf{G}}
&:=&\mathbf{E}\left[\left.\uti^{\prime}(x+\varphi\mal S_{T})
\right|\mathcal{G}_{t}\right]\nonumber \\
& = & \mathbf{E}\left[\left.p\exp\left(-px-\int_{0}^{T}\frac{b(Y_{s})}
{\sigma(Y_{s})}dW_{s}-\int_{0}^{T}\Big(\frac{b(Y_{s})}{\sigma(Y_{s})}\Big)^{2}ds\right)
\right|\mathcal{G}_{t}\right]\nonumber \\
 & = & \mathbf{E}\left[\left.pe^{-px}\widetilde{Z}_{T}
 \overline{Z}_{T}\right|\mathcal{G}_{t}\right]\nonumber \\
 & = & \underbrace{pe^{-px}\widetilde{Z}_{T}}_{\mathcal{G}_{0}\textrm{-measurable}}
 \overline{Z}_{t}\label{eq:verallg. BS-Z_G}
\end{eqnarray}
for any $t\in[0,T]$.
In particular, $\mathbf{E}[Z_{T}^{\mathbf{G}}]<\infty$.
The normalised $\mathbf{G}$-martingale $Z^\mathbf{G}$ is the $\mathbf{G}$-density process
of the probability measure $\mathbf{Q}$ with density
\[
\frac{d\mathbf{Q}}{d\mathbf{P}}:=\frac{Z_{T}^{\mathbf{G}}}
{\mathbf{E}\left[Z_{T}^{\mathbf{G}}\right]}.
\]

\emph{Step 2:}
Let $\mathbf{\overline{Q}}$ be the probability measure with density
process $\overline{Z}$. By Girsanov's theorem,
\[
W^{\mathbf{\overline{Q}}}
:=W+\int_{0}^{\cdot}\frac{b(Y_{t})}{\sigma(Y_{t})}dt
\]
is a standard Brownian motion under measure $\mathbf{\overline{Q}}$ relative
to both $\mathbf{F}$ and $\mathbf{G}$. Since ${b(Y)}/{\sigma(Y)}$
is bounded, 
\[
 \varphi\mal S=\frac{1}{p}\int_{0}^{\cdot}\frac{b(Y_{t})}
{\sigma(Y_{t})}dW_{t}^{\mathbf{\overline{Q}}}
\]
is a $\mathbf{\overline{Q}}$-martingale with respect to $\mathbf{G}$.
Moreover, $S$ is a $\mathbf{\overline{Q}}$-local martingale relative
to both $\mathbf{F}$ and $\mathbf{G}$ because $dS_{t}=S_{t}
\sigma(Y_{t})dW_{t}^{\mathbf{\overline{Q}}}$, cf.\ \cite[Theorem IV.33]{protter.04}. 

Let $\psi$ be an admissible
strategy in the sense of \cite[Definition 1.2]{schachermayer.99}, i.e.\
$\psi$ is an 
$S$-integrable process such that the related wealth process is uniformly
bounded from below. Note that $\psi\mal S$ is a $\mathbf{\overline{Q}}$-local
martingale which is bounded from below and hence a $\mathbf{\overline{Q}}$-supermartingale. 
By the generalized Bayes' formula and in view of (\ref{eq:verallg. BS-Z_G}),
$\varphi\mal S$ is
a $\mathbf{Q}$-martingale and $\psi\mal S$ is a $\mathbf{Q}$-supermartingale,
both with respect to filtration $\mathbf{G}$.
Hence, by concavity of $\uti$, we have
\begin{eqnarray*}
 \lefteqn{\mathbf{E}[\uti(x+\psi\mal S_{T})]}\\
 & \leq & \mathbf{E}\left[\uti(x+\varphi\mal S_{T})\right]+\mathbf{E}\left[\uti^{\prime}
 (x+\varphi\mal S_{T})(\psi\mal S_{T}-\varphi\mal S_{T})\right]\\
 & = & \mathbf{E}\left[\uti(x+\varphi\mal S_{T})\right]+
 \mathbf{E}\left[\uti^{\prime}(x+\varphi\mal S_{T})\right]
 \underbrace{\mathbf{E}_{\mathbf{Q}}\left[\psi\mal S_{T}-\varphi\mal S_{T}\right]}_{\leq0}\\
 & \leq & \mathbf{E}\left[\uti(x+\varphi\mal S_{T})\right].
\end{eqnarray*}

\emph{Step 3:}
We show that $\uti(x+\varphi\mal S_{T})$ lies in
the $L^{1}$-closure of the set
\[
\{ \uti(x+\psi\mal S_{T}):\psi\textrm{ is admissible}\} .
\]
Indeed, letting
\[
\tau_{n}:=\inf\{ t\in[0,T]: x+\varphi\mal S_{t}<-n\} ,
\]
we can approximate $\varphi$ by the sequence $(\varphi^{(n)})_{n\in\mathbb{N}}$
defined as $\varphi^{(n)}:=\varphi\mathbf{1}_{\auf0,\tau_{n}\zu}$,
which fulfills the admissibility requirement in \cite[Definition 1.2]{schachermayer.99}.
Using Cauchy-Schwarz' inequality, we obtain
\begin{eqnarray*}
\lefteqn{\mathbf{E}\left[\sup_{n\in\mathbb{N}}\left|\uti(x+\varphi^{(n)}\mal S_{T})\right|\right]}\\
 & = & \mathbf{E}\left[\sup_{n\in\mathbb{N}}\exp\big(-p(x+\varphi^{(n)}\mal S_{T})\big)\right]\\
 & \leq & \mathbf{E}\left[\sup_{t\in[0,T]}\exp\big(-p(x+\varphi\mal S_{t})\big)\right]\\
 & = & e^{-px}\mathbf{E}\left[\sup_{t\in[0,T]}\exp\left(-\int_{0}^{t}\frac{b(Y_{s})}
 {\sigma(Y_{s})}dW_{s}-\frac{1}{4}\int_{0}^{t}
 \Big(\frac{2b(Y_{s})}{\sigma(Y_{s})}\Big)^{2}ds\right)\right]\\
 & \leq & e^{-px}\sqrt{\mathbf{E}\Bigg[\sup_{t\in[0,T]}\underbrace{\exp\left(-\int_{0}^{t}
 \frac{2b(Y_{s})}{\sigma(Y_{s})}dW_{s}-\frac{1}{2}\int_{0}^{t}\Big(\frac{2b(Y_{s})}
 {\sigma(Y_{s})}\Big)^{2}ds\right)}_{=:M_{t}}\Bigg]}.
\end{eqnarray*}
By boundedness of $({2b(Y_{t})}/{\sigma(Y_{t})})_{t\in[0,T]}$,
the process $(M_{t})_{t\in[0,T]}$ is an $L^{2}$-martingale,
which implies 
\[
\sup_{n\in\mathbb{N}}\uti(x+\varphi^{(n)}\mal S_{T})\in L^{1}(\mathbf{P})
\]
by Doob's quadratic inequality.
Dominated convergence yields
\[
\left\Vert \uti(x+\varphi^{(n)}\mal S_{T})-\uti(x+\varphi\mal S_{T})\right\Vert _{L^{1}(\mathbf{P})}\to0
\quad\textrm{as}\quad n\to\infty.
\]

\emph{Step 4:}
By Steps 2 and 3 the payoff $x+\varphi\mal S_{T}$ is
optimal in the sense of \cite[Theorem 2.2(iii)]{schachermayer.99}, which
implies that $\mathbf{Q}$ is the dual optimizer, cf.\ \cite[Equation (42)]{schachermayer.99}. 
Moreover, we have shown in Step 2 that
$\varphi\mal S$ is a $\mathbf{Q}$-martingale with respect
to filtration $\mathbf{G}$ and hence $\mathbf{F}$ as well, which
yields that $\varphi$ is the optimal strategy in the sense of
\cite[Theorem 2.2(iv)]{schachermayer.99}.
\end{prf}

\end{appendix}

\bibliography{shenbib}
\end{document}